\documentclass[reprint,
amsmath,amssymb,
aps,
]{revtex4-2}

\usepackage{graphicx}
\usepackage{dcolumn}
\usepackage{bm}
\usepackage{color}
\usepackage{color}
\usepackage[usenames,dvipsnames]{xcolor}
\usepackage[citecolor=blue, colorlinks=true, urlcolor=blue, linkcolor=blue]{hyperref}

\newcommand{\req}[1]{Eq.$\,$(#1)}
\newcommand{\rfig}[1]{Fig.$\,$(#1)}
\newcommand{\RomanDigit}[1]{\MakeUppercase{\romannumeral #1}}

\begin{document}

\preprint{APS/123-QED}

\title{Fractionalized Prethermalization in the One-Dimensional Hubbard Model} 

\author{Anton Romen\textsuperscript{1,2}}
\author{Johannes Knolle\textsuperscript{1,2,3}}
\author{Michael Knap\textsuperscript{1,2}}
\affiliation{\textsuperscript{\normalfont 1}Technical University of Munich, TUM School of Natural Sciences, Physics Department, 85748 Garching, Germany
\\ \textsuperscript{\normalfont 2}Munich Center for Quantum Science and Technology (MCQST), Schellingstraße. 4, 80799 München, Germany
\\ \textsuperscript{\normalfont 3}Blackett Laboratory, Imperial College London, London SW7 2AZ, United Kingdom}

\date{\today}

\begin{abstract}
 Prethermalization phenomena in driven systems are generally understood via a local Floquet Hamiltonian obtained from a high-frequency expansion. Remarkably, recently it has been shown that a driven Kitaev spin liquid with fractionalized excitations can realize a quasi-stationary state that is not captured by this paradigm. Instead distinct types of fractionalized  excitations are characterized by vastly different temperatures---a phenomenon dubbed \textit{fractionalized prethermalization}. In our work, we analyze fractionalized prethermalization in a driven one-dimensional Hubbard model at strong coupling which hosts spin-charge fractionalization. At intermediate frequencies quasi-steady states emerge which are characterized by a low spin and high charge temperature with lifetimes set by two competing processes:  the lifetime of the quasiparticles determined by Fermi's Golden rule and the exponentially long lifetime of a Floquet prethermal plateau. We classify drives into three categories, each giving rise to distinct (fractional) prethermalization dynamics. Resorting to a time-dependent variant of the Schrieffer-Wolff transformation, we systematically analyze how these drive categories are linked to the underlying driven Hubbard model, thereby providing a general understanding of the emergent thermalization dynamics. We discuss routes towards an experimental realization of this phenomenon in quantum simulation platforms.
\end{abstract}

\maketitle

\section{\label{sec:intro}Introduction}

\begin{figure}
    \centering
    \includegraphics{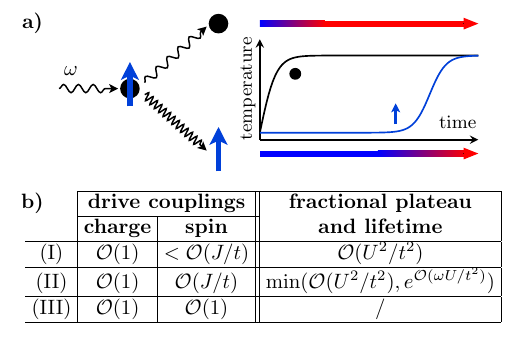}
    \caption{\textbf{Fractionalized prethermalization arising from spin-charge separation in the one-dimensional Hubbard model:} a) Fractionalization of fermions causes a periodic drive with frequency $\omega$ to typically couple asymmetrically to the fast charge quasi-particles of scale $t$ and slow spin quasi-particles of scale $J = 4t^2/U$, where $U$ is the Hubbard interaction. The latter experiences a higher relative frequency due to the smaller intrinsic energy scale at strong coupling, $J\ll t$. This results in a fractional prethermal plateau characterized by fast heating of the charge sector and a low effective temperature prethermal state in the spin sector. b) The lifetime of the resulting plateau is determined by the coupling strength of the drive to the spin quasi-particles. Three classes should be distinguished: (I) For drives that do not directly couple to the spinons, a  fractionalized prethermal regime exists and its lifetime is limited only by the quasi-particle lifetime $\mathcal{O} (U^2/t^2)$. (II) For a weak coupling of the drive to the spin degrees of freedom on the order of $J$, the lifetime of the fractionalized prethermal plateau is determined by the competition between the quasi-particle lifetime $\mathcal{O} (U^2/t^2)$ and the exponential lifetime of a prethermal plateau in the spin sector $\mathcal{O} (e^{\omega U/t^2})$. (III) When both charge and spin is strongly driven the system heats rapidly. A fractional plateau is not observable. }
    \label{fig:basics-sketch1}
\end{figure}

Strong periodic modulation has been established as a powerful route to engineer effective Hamiltonians with properties that are distinct from their undriven counterparts~\cite{oka2009photovoltaic,lindner2011floquet, goldman2014periodically,bukov2015prethermal,eckardt2017colloquium,cooper2019topological,oka2019floquet,rudner2020band}. With that approach dynamical phases of matter may be realized that cannot even exist in thermal equilibrium. Those include Floquet anomalous Hall phases in non-interacting systems which host chiral edge states even though the bulk bands have zero Chern number~\cite{kitagawa2010topological,rudner2013anomalous,roy2017periodic, wintersperger2020realization, braun2024real}, as well as phases of matter with intrinsic Floquet topological order~\cite{po2017radical,fulga2019topology}, which has been recently demonstrated on a quantum processor~\cite{will2025probing}. Although these phases of matter are in principle robust toward some perturbations, interacting many-body systems generically absorb energy from the drive and heat toward a featureless infinite temperature state. A common strategy to avoid this challenge is to resort to driving frequencies that are high compared to internal energy scales, giving rise to an exponential suppression of heating rates~\cite{bukov2015universal, kuwahara2016floquet, abanin2015exponentially, canovi2016stroboscopic, mori2016rigorous, weidinger2017floquet, abanin2017rigorous, mori2018floquet}. These quasi-steady states can be stable for exponentially long times and are referred to as prethermal states, that have been explored in various settings, see e.g. Refs.~\cite{mallayya2019heating, mallayya2019prethermalization,kuhlenkamp2020periodically,fleckenstein2021prethermalization, fleckenstein2021thermalization, oDea2024prethermal, zhao2021random, pawowski2024long, mcRoberts2023prethermalization}. They are described by an effective low-temperature Gibbs ensemble of a local effective Hamiltonian~\cite{kuhlenkamp2020periodically}.

The effective Hamiltonian, describing the prethermalization dynamics, can be computed from a high-frequency expansion. However, this scenario has been recently challenged in a driven two-dimensional Kitaev honeycomb model, with fractionalized flux and matter excitations, where it has been shown that the prethermal regime hosts different temperatures for the fractionalized excitations even though the physical spin degree of freedom is periodically modulated~\cite{jin2023fractionalized}. The resulting prethermal state cannot be described by an effective Hamiltonian obtained from a high-frequency expansion at a single temperature. Instead, it is characterized by vastly different effective temperatures of the fractionalized excitations. This phenomenon was therefore termed \textit{fractionalized prethermalization}. 
The Kitaev honeycomb model is, however, quite special as the full Hilbert space separates into flux and matter sectors at all energies~\cite{kitaev2006anyons}. It therefore remains to be seen whether other systems with fractionalized excitations host similar fractionalized prethermal quasi-steady states. Furthermore, it is unknown how robust the phenomenon is toward changing the family of drive protocols and what the influence of a finite lifetime of quasiparticles could be on the quasi-steady state.

In this work, we address these pertinent questions by investigating the one-dimensional hole-doped Hubbard model at strong coupling as a paradigmatic example of fractionalized spin and charge excitations~\cite{meden1992spectral,voit1993charge,giamarchi2003quantum}. Furthermore, this model is naturally realized in experiments, for example with ultracold atoms~\cite{viJayan2020time} where different drive protocols can be engineered as well~\cite{aidelsburger2013realization, miyake2013realizing, jotzu2014experimental, bordia2017periodically, wintersperger2020realization, rubioAbadal2020floquet}, rendering it an ideal system for exploring fractionalized prethermalization.
We find a long-lived fractionalized prethermal plateau when the model is driven at frequencies that are on the order of the hole hopping scale but large compared to the spin-exchange energy scale. The fractionalized prethermal regime is characterized by fast heating of the charge sector, while energy absorption in the spin sector is smaller by orders of magnitude. Using a time-dependent variant of the Schrieffer-Wolff transformation we systematically analyze the consequences of different drives on the low-energy effective theory and find three distinct regimes, see \rfig{\ref{fig:basics-sketch1}}: Class (I) in which the drive  couples only to the charge sector but not to the spin, leads to a fractional prethermal plateau with Fermi Golden rule (FGR) lifetime determined by
the quasiparticle lifetime. Class (II) is a drive that couples asymmetrically to the spin and charge sectors, resulting in a fractional prethermal plateau that is limited by the smaller of either a  lifetime exponential in frequency or by the quasi-particle lifetime. Class (III) couples both quasiparticles strongly and leads to a breakdown of spin-charge separation, which in turn causes rapid global thermalization.

We organize our work  as follows: In section \RomanDigit{2} we introduce the model and background on spin-charge fractionalization in one dimension. Our results on fractionalized prethermalization for drives of Class (I) and Class (II) in the hole-doped Hubbard model are illustrated in Section \RomanDigit{3}; Class (III) is analyzed in the appendix. In section \RomanDigit{4} we systematically derive the  structures of the effective theories starting from a one-dimensional Hubbard model. We introduce prospects for an experimental realization in section \RomanDigit{5} and present our outlook in \RomanDigit{6}. Technical details are relegated to the appendices.
\vspace{5mm}

\section{Fractionalization of fermions in one dimension}
\label{sec:squeezed-space}

The starting point of our discussion is the one-dimensional Hubbard model of strongly correlated lattice fermions
\begin{eqnarray}
    H_\mathrm{Hub} = -t\sum_{j,\alpha} (c^\dagger_{j,\alpha}c_{j+1, \alpha} + \mathrm{H.c.}) + \frac{U}{2}\sum_j n_j(n_j-1). \hspace{5mm}         
\label{eq:Hubbard-model}
\end{eqnarray}
Here, $t$ denotes the hopping amplitude, $U$ the onsite interaction strength and $n_j = \sum_\alpha c_{j,\alpha}^\dagger c_{j,\alpha}$ the number operator on site $j$. In the strongly repulsive limit $U \gg t$ and below half filling the low-energy effective theory up to order $\mathcal{O}(t^2/U)$ is governed by the tJ-model 
\begin{widetext}
\begin{eqnarray}
    H_{tJ} = \mathcal{P}_0 \bigg[ -t \sum_{j,\alpha=\uparrow, \downarrow}(c^\dagger_{j,\alpha}c_{j+1, \alpha} + \mathrm{H.c.}) \bigg]\mathcal{P}_0 + J \sum_j \bigg( \mathbf{S}_{j}\mathbf{S}_{j+1} - \frac{n_{j}n_{j+1}}{4} \bigg) 
\label{eq:tJ-model}
\end{eqnarray}
\end{widetext}
see e.g. Ref. \cite{auerbach1994interacting}. Here, $\mathcal{P}_0$ projects out doubly occupied sites, the spin operators are given by $\mathbf{S}_j=\sum_{\alpha,\alpha'} c_{j,\alpha}^\dagger \mathbf{\sigma}_{\alpha,\alpha'} c_{j,\alpha'}$ where $\mathbf{\sigma}$ denotes the Pauli matrices and $J=4t^2/U\ll t$ denotes the strength of the effective spin interaction.
We consider the tJ-model in the the weakly hole-doped regime and neglect an additional small term $\mathcal{O}(J/8)$ arising in the derivation of the tJ-model \cite{auerbach1994interacting}. Remarkably, in this model, charge defects that proliferate through the chain are effectively decoupled from the spins resulting in spin-charge separation~\cite{meden1992spectral,voit1993charge,giamarchi2003quantum}. The resulting equilibrium properties were recently studied in \cite{bohrdt2018angle}.

To gain an intuitive picture, one first notes that the charge defects or holes are only weakly coupled to the spins and can therefore move almost freely through the chain. Additionally, the dynamics of the holes create defects in the spin chain resulting in excitations of the spin chain. In summary, this allows one to distinguish two quasiparticles: the fast \textit{chargon} associated with the movement of charge defects on a timescale set by the hopping amplitude $t$ and a slow \textit{spinon} describing the excitations in the spin chain and moving on a timescale set by $J$. The spectral building principle \cite{bannister2000spectral} can then be used to determine at which energies and momenta excitations can occur.

The observation that charge and spin excitations behave as nearly independent particles motivates a formal description using a slave particle formulation, see \cite{bohrdt2018angle} and Appendix \ref{A:squeezed-space} for a detailed discussion. The key idea is to introduce two partons, the bosonic chargon $h_j$ accounting for the charge defects in the chain and fermionic spinon $f_{j,\alpha}$ describing the spins. The original fermionic operators are then expressed as $c_{j,\alpha} \rightarrow h_j^\dagger f_{j,\alpha}$. Subsequently, only spins $\tilde{f}_{j,\alpha}$ on occupied sites are considered which effectively maps the chargons onto the bonds of the chain. Fock states in the transformed representation then consist of $N$ spinons residing on the sites of the squeezed space and $L-N$ chargons on the links between, where $L$ is the system size.

\begin{figure}
    \centering
    \includegraphics{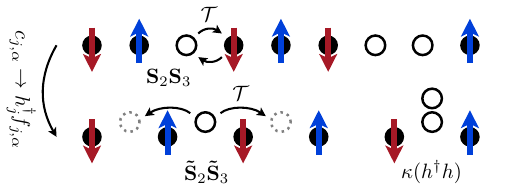}
    \caption{\textbf{a) Squeezed space formalism:} Spin-charge separation in the 1D tJ-model is understood by a parton construction, where holes reside on the bonds of a squeezed spin chain. Multiple adjacent holes are mapped to the same bond. Within the parton description, the kinetic energy $\mathcal{T}$ then corresponds to bosonic chargons hopping between links of the chain and Heisenberg interactions $\tilde S_i \tilde S_{i+1}$ between fermionic spinons are turned off when chargons reside on the bond in between. The density interactions $\kappa(h^\dagger h)$ are related to the number of holes on each bond, see main text for details.}
    \label{fig:basics-sketch2}
\end{figure}

The tJ-model from \req{\ref{eq:tJ-model}} is thus transformed as follows: First, the constrained hopping of particles on the chain corresponds to chargons freely hopping on the links. The spin interaction maps to Heisenberg interactions of spinons unless a chargon is present on the link and, $k\geq 1$ chargons residing on a link switches off $k+1$ density interactions. This results in the representation
\begin{eqnarray}
    \begin{aligned}
    H_{tJ} &= \mathcal{T} + \mathcal{S} +\mathcal{D} \\
    \mathcal{T} &= -t\sum_{j} (\tilde{h}^\dagger_{j+1/2} \tilde{h}_{j+3/2} +\mathrm{H.c.}) \\
    \mathcal{S} &= J \sum_j \tilde{\mathbf{S}}_{j}\tilde{\mathbf{S}}_{j+1}\cdot(1-\Theta(\tilde{h}_{j+1/2}^\dagger \tilde{h}_{j+1/2})) \\
    \mathcal{D} &= \frac{1}{4} \bigg[ L - \sum_j \kappa(\tilde{h}_{j+1/2}^\dagger \tilde{h}_{j+1/2}) \bigg ] 
    \end{aligned}
    \label{eq:tJ-model-squeezed-space}
\end{eqnarray}
as derived in Appendix \ref{A:squeezed-space} and shown schematically in \rfig{\ref{fig:basics-sketch2}}. We use the step function $\Theta(x)$ and define $\kappa(x) = \Theta(x) \cdot (x+1)$. Additionally, to emphasize that holes are residing on the links, we explicitly map $h_j \rightarrow \tilde{h}_{j+1/2}$.

From \req{\ref{eq:tJ-model-squeezed-space}} spin-charge fractionalization is immediately apparent. Both the kinetic energy $\mathcal{T}$ and density interactions $\mathcal{D}$ depend only on the chargon and the spin interaction $\mathcal{S}$ can be separated into a Heisenberg term $H_\text{Heis}$ coupling only to  spinons and an interaction term $H_\mathrm{int}$ with  energy scale $J$
\begin{eqnarray}
    \mathcal{S} &=& H_\text{Heis} + H_\mathrm{int} \nonumber\\&=& -J \sum_j \tilde{\mathbf{S}}_{j}\tilde{\mathbf{S}}_{j+1}-J \sum_j \tilde{\mathbf{S}}_{j}\tilde{\mathbf{S}}_{j+1} \Theta(\tilde{h}_{j+1/2}^\dagger \tilde{h}_{j+1/2}). \nonumber
\end{eqnarray}
 To make the general structure apparent, let us rewrite
\begin{eqnarray}
    H_{tJ} = \mathcal{T} + \mathcal{D} + \mathcal{S} \equiv H_0 + H_{\text{int}}.
    \label{eq:tJ-model-description}
\end{eqnarray}
Here, $H_0$ does not couple chargons and spinons and corresponds to the assumption of independent quasiparticles. A finite lifetime of quasiparticles is induced by the coupling term. In the strongly correlated limit $J \ll t$ we can view $H_\text{int}$ as a small perturbation to the non-interacting $H_0$ and employ Fermi's golden rule which predicts 
\begin{eqnarray}
    t\tau_0 \sim \frac{t^2}{J^2} \sim \frac{U^2}{t^2}
\label{eq:quasiparticle lifetime}
\end{eqnarray}
for the lifetime $\tau_0$ of the quasiparticles. This result in particular implies that considering the chargon and spinon independently is a valid approximation up to times $\mathcal{O}(\tau_0)$. 

\section{Fractionalized prethermalization}
\label{sec:fractionalized-prethermalization}

Before we introduce the periodically driven system, let us briefly specify our notation, where $t$ denotes hopping amplitude, $\tau$ real time, $\beta$ inverse temperature and $\omega$ the frequency of the drive.

Our main focus are the stroboscopic Floquet dynamics of a periodically driven tJ-model
\begin{eqnarray}
    H(\tau) = H_{tJ}+gf(\tau)\hat{V}.
    \label{eq:tJ-model-with-drive}
\end{eqnarray}
Here, $\hat{V}$ is a drive of moderate strength $g \sim t$ and we require that $f(\tau+T)=f(\tau)$ is of order $\mathcal{O}(1)$ and periodic with period $T=\frac{2\pi}{\omega}$. 

Generally, the drive $\hat{V}$ will induce strong couplings between the two partons and is thus expected to break spin-charge separation of the tJ-model. For small to intermediate driving frequencies $\omega \sim t$ the system is then expected to rapidly heat up and thermalize and the  eigenstate thermalization hypothesis predicts that the long time steady state is locally indistinguishable from an infinite temperature state.  Conversely, in the high-frequency limit $\omega \gg t$, energy from the drive can only be absorbed via perturbative processes of order $n=\omega/t$. The corresponding matrix elements are then typically of order $\mathcal{O}(e^{-n})$, indicating exponential suppression of heating. Formally, this behavior is obtained from a high frequency expansion $H_\text{eff} = H_{tJ}+\mathcal{O}(t/\omega)$ \cite{bukov2015universal,kuwahara2016floquet,abanin2015exponentially}, which describes the dynamics up to times $\tau \sim e^\omega$. These prethermalization arguments predict a long-lived quasi-steady state, termed prethermal state, that is described by a low temperature Gibbs ensemble of the effective Hamiltonian $H_\text{eff}$ \cite{bukov2015prethermal}.

Let us now try to gain some intuition how the dynamics created by \req{\ref{eq:tJ-model-with-drive}} can lead to unconventional thermalization behavior, particularly fractionalized prethermalization. According to Ref. \cite{jin2023fractionalized}, the two key factors leading to fractional prethermal plateaus are 1) a drive that couples asymmetrically to the quasiparticles and 2) quasiparticles residing at different local energy scales.

First, consider a drive that couples only to the chargon. For low to intermediate drive frequencies, the charge sector is then expected to thermalize on a short timescale, but due to the separation of scales, \req{\ref{eq:quasiparticle lifetime}}, the spinon can only absorb energy via the weak static coupling. Consequently, heating of the spinon is drastically suppressed when the quasiparticle lifetime is much longer than the heating time of the chargon. In that regime the system realizes a fractional prethermal plateau of lifetime governed by the quasiparticle lifetime, where the heating time of the spin sector is much longer than the one of the charge sector. Note that a similar situation holds in general for large $U$ as long as drive induced couplings between the chargon and spinon are subleading compared to the static coupling in \req{\ref{eq:tJ-model-description}}.

A second scenario leading to a fractional prethermal plateau results from a drive that couples strongly to the chargon while coupling the chargon and spinon weakly on the order of $J$, such that the dynamic coupling is on the order of the static interaction in \req{\ref{eq:tJ-model-description}}. Again, the strong coupling to the chargon results in rapid thermalization of the charge sector at small to intermediate frequencies. However, additionally such a drive couples significantly to the spinon. Due to the small energy scale, the spinon is however driven at much higher relative driving frequency than the chargon. One may then employ prethermalization arguments to predict that heating rates of the spinon are exponentially suppressed in frequency with the quasi-steady state in the spin sector given by a low temperature Gibbs ensemble of an effective spinon Hamiltonian. The lifetime of this prethermal spin state is determined by a competition between the quasiparticle lifetime and exponential lifetime of the prethermal plateau where the smaller one of the two is dominating.

We show in Sec. \ref{sec:Floquet-expansion-Hubbard} that the two cases introduced above can arise from the dynamics of the underlying driven Hubbard model: A drive that couples both quasiparticles strongly causes a breakdown of spin-charge fractionalization resulting in generic thermalization behavior. By contrast, drives that couple only to the chargon or asymmetrically to the chargon and spinon lead to a fractional prethermal plateau. Independent of the lifetime of this plateau, a distinguishing feature of the quasi-steady state is the emergence of distinct effective temperatures of the chargon and spinon, respectively, as summarized in \rfig{\ref{fig:basics-sketch1}}.

In the following subsections we investigate the above considerations numerically by studying the stroboscopic dynamics of the driven tJ-model in \req{\ref{eq:tJ-model-with-drive}} using exact diagonalization techniques. We consider two cases: a drive that couples only to the chargon and probes relaxation in the spin sector due to the static coupling and a drive that couples to both quasiparticles and probes the exponential lifetime of the fractional prethermal plateau as function of frequency.

For numerical simplicity we choose a piecewise constant driving protocol such that the evolution operator over one period acquires the simple form
\begin{eqnarray}
    U(T) = e^{-i\frac{T}{2}(H_{tJ}-g\hat{V})} e^{-i\frac{T}{2}(H_{tJ}+g\hat{V})} = e^{-iTH_\text{eff}}
\end{eqnarray}
with $H_\text{eff}$ the effective Floquet Hamiltonian. For large frequencies, the Magnus expansion predicts $H_\text{eff} = H_{tJ}-\frac{iT}{4}[H_{tJ},\hat{V}]$ recovering the tJ-model as effective Hamiltonian in the infinite frequency limit, $T\to0$. Throughout this work we are instead interested in moderate frequencies where the leading-order Magnus expansion is not a good approximation.

The heating dynamics of the charge and spin sector can be captured by constructing suitable observables: Since the charge sector corresponds to almost free particles, we expect the dynamics to be well captured by the kinetic energy
\begin{eqnarray}
    K = \sum_{j,\alpha} (c_{j,\alpha}^\dagger c_{j+1,\alpha} +\text{H.c.}).
    \label{eq:observable-kin}
\end{eqnarray}
The dynamics in the spin sector are governed by a Heisenberg Hamiltonian and exploiting SU(2) symmetry, we can use the coupling in z-direction
\begin{eqnarray}
    Z = \sum_{j} S^z_j S^z_{j+1}
    \label{eq:observable-zz}
\end{eqnarray}
to capture the energy of the spins. To simplify the analysis, we subtract the infinite temperature expectation value of the observables, e.g.,
\begin{eqnarray}
    \tilde{Z} \equiv Z - \text{Tr}(Z)/\text{Tr}(\mathcal{I})
\end{eqnarray}
such that $\langle \tilde{Z}(\beta \rightarrow 0) \rangle \rightarrow 0$.

In our simulations, we first prepare a typical thermal state $|\psi(\tau=0)\rangle$ at some low temperature $1/\beta_0$ using the method of typicality \cite{bartsch2009dynamical,reimann2018dynamical,sugiura2012thermal}. 
Starting from a set of random complex vectors ${|r_j\rangle}_{j=1,...,M}$, a typical thermal state can then be prepared by iteratively computing $|r_j(n)\rangle = e^{-\beta_\delta/2H}|r_j(n-1)\rangle$ with some small stepsize $\beta_\delta$ until the desired temperature is reached and subsequently averaging over the obtained states. After that, we evaluate the expectation values $\langle K(\tau) \rangle$ and $\langle Z(\tau) \rangle$ with respect to the time-evolved state $|\psi(\tau = nT)\rangle = [U(T)]^n|\psi(0)\rangle$ for different times up to $t\tau=10^5$ using Krylov subspace methods of Ref.~\cite{weinberg2019quspin}. We explicitly enforce translational symmetry and work in the zero total momentum sector. By further exploiting the $U(1)$ charge and spin symmetries as well, we can simulate systems with up to $L=22$ sites and $2$, $3$ or $4$ holes in the system. Nonetheless, we encounter considerable finite size effects due to the product structure of the Hilbert space $\mathcal{H} = \mathcal{H}_h \otimes \mathcal{H}_s$ in terms of the charge $\mathcal{H}_h$ and spin sector $\mathcal{H}_s$. As a consequence, the resulting dynamics are sometimes not fully ergodic and not all observables relax to their true infinite temperature expectation value at large times. A standard approach in the literature is to introduce a small randomness $T \rightarrow T(1+\delta)$ in each drive period, which is expected to restore ergodicity \cite{fleckenstein2021prethermalization,fleckenstein2021thermalization}. However, this approach introduces an additional uncontrolled relaxation channel \cite{sieberer2018statistical} and causes misleading and inconclusive scaling predictions for the lifetime of the fractional plateau, see Appendix \ref{A:additional-data} for details. Therefore, in our analysis, we instead address the problem of incomplete thermalization by rescaling the spin energy with respect to the late time value instead of the true infinite temperature value to extract the scaling predictions.

We now discuss the results of the two different drive protocols introduced above. 

\subsection{Class (I): Driving Charge -- Staggered potential}
\label{sec:frac-pretherm-potential}

\begin{figure}[h]
\includegraphics{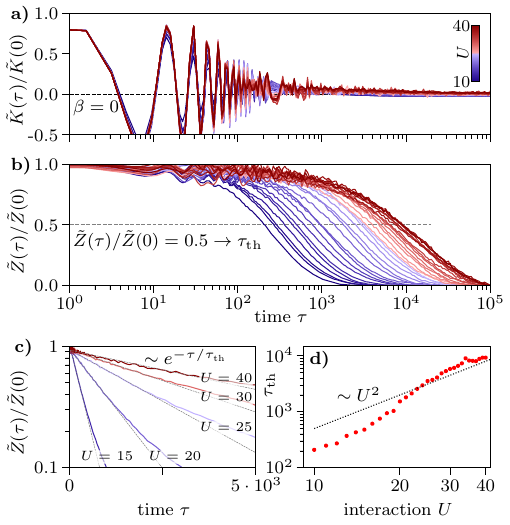}
\caption{\textbf{Fractionalized prethermalization with Fermi Golden Rule  lifetime.} The heating dynamics of a driven tJ-model shows a two-step structure. (a) Due to the strong coupling of the chargon with the drive, the charge sector heats up rapidly as indicated by the kinetic energy, which fluctuates around zero after an initial relaxation time. (b) In contrast, energy cannot be absorbed into the spin sector effectively, resulting in a long lived quasi-steady state. This state is captured by the spin-spin coupling $\tilde{Z}$ which retains most of its initial value for times up to $t\tau \sim 10^4$. (c,d) The lifetime of the quasi-steady spin state is determined by the breakdown of quasiparticles, resulting in a Fermi Golden Rule scaling of the lifetime $\tau_\mathrm{th} \sim U^2$ at large $U$ and an exponential decay $\sim e^{-\tau/\tau_\mathrm{th}}$ of the spin energy.}
\label{fig:staggered-pot-drive}
\end{figure}

We first drive the model with a staggered potential of the form
\begin{eqnarray}
    \hat{V}_h = \sum_j (-1)^j n_j.
    \label{eq:staggered-potential-drive}
\end{eqnarray}
which is spin independent as it couples only to the total particle density. Thus, within the squeezed space formulation, we expect that this drive couples only to the chargon; see Appendix \ref{A:squeezed-space}. In our simulations, we keep hopping amplitude $t=1$, frequency $\omega=4t$ and drive strength $g=0.324t$ fixed while varying $U=4t^2/J \in [10,40]$. As shown in \rfig{\ref{fig:staggered-pot-drive}} (a) and (b), dynamics of the kinetic energy of holes and the energy of spins are radically different, giving rise to a fractionalized perthermal plateau. The kinetic energy starts to fluctuate around a value close to zero on a timescale $t\tau \sim 10^2$ independently of $U$, indicating a high temperature state in the charge sector.  By contrast, relaxation times of the spin sector are larger by orders of magnitude and show a strong dependence on $U$.

\begin{figure}[h]
\includegraphics{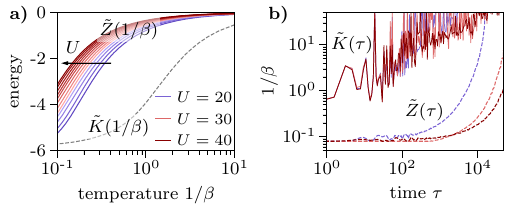}
\caption{\textbf{Absence of conventional prethermalization.} (a) Kinetic energy $\tilde{K}$ and spin correlations $\tilde{Z}$ in thermal equilibrium, representing hole and spin energies, respectively. (b) During the heating process, the effective temperature of the spin and charge sector differ by multiple orders of magnitude over a large temporal regime of $10^2 \lesssim t\tau \lesssim 10^4$ leading to an unconventional fractional prethermal quasi-steady state. Simulation parameters are the same as in \rfig{\ref{fig:staggered-pot-drive}}.  }
\label{fig:staggered-pot-temperatures}
\end{figure}

Since the drive does not couple to the spin directly, energy can be absorbed into the spin sector only via the static interaction from \req{\ref{eq:tJ-model-description}}. We therefore expect that heating in the spin sector is determined by the lifetime of the spinon given in \req{\ref{eq:quasiparticle lifetime}}. Consequently, we expect that the lifetime $\tau_\mathrm{th}$ of the prethermal plateau also follows a FGR $\tau_\mathrm{th} \sim U^2$ for large enough $U$. To gain a quantitative numerical estimate, we first define the lifetime $\tau_\mathrm{th}$ as the time where the spin coupling $\tilde{Z}(\tau_c)/\tilde{Z}(0)=0.5$. Indeed, up to some fluctuations, the thus-obtained lifetime $\tau_\mathrm{th} \sim U^2$ shows the predicted FGR scaling for large $U$. To reduce fluctuations, we make use of a result established for weakly periodically driven systems \cite{mallayya2019heating}. The key idea is to consider the absorption of energy over one period in linear response theory, which leads to a FGR for the heating rate $\Gamma = \dot E(\tau)/|E(\infty)-E(\tau)|$ of the system. In turn, the above implies that the energy is exponentially decaying in time, reminiscent of the survival probability of quasiparticles in standard FGR calculations. We fit this exponential decay of the spin coupling to obtain a more stable numerical estimate of the lifetime, \rfig{\ref{fig:staggered-pot-drive}}(c), and recover again the predicted FGR scaling in agreement with the previously obtained values as shown in \rfig{\ref{fig:staggered-pot-drive}}(d). 

So far, we have characterized the distinct heating dynamics of charge and spin degrees of freedom by inspecting their relaxation dynamics. We now further analyze the fractionalized prethermal state by extracting the effective temperatures of the quasiparticles.  To do so, we first compute the thermal expectation values of the charge kinetic energy $\tilde{K}(\beta)$ and the spin energy $\tilde{Z}(\beta)$ and use them to determine the effective temperature in both sectors as function of time. We find that in the fractionalized prethermal regime these temperatures differ by two orders of magnitude, suggesting that the established quasi-steady state is not a conventional prethermal state; see \rfig{\ref{fig:staggered-pot-temperatures}}. In particular, we emphasize that the resulting quasi-steady state cannot be described by an effective low-temperature Gibbs ensemble $\rho \nsim e^{-\beta_\text{eff}H_\text{eff}}$, obtained from the effective Hamiltonian that is constructed via the optimal-order Floquet-Magnus expansion. Explicit numerical results for the underlying Hubbard model are shown below in \rfig{\ref{fig:staggered-pot-Hubbard}}(c). 

\subsection{Class (II): Driving charge and spin -- Staggered hopping and Heisenberg term}

\begin{figure}
\includegraphics{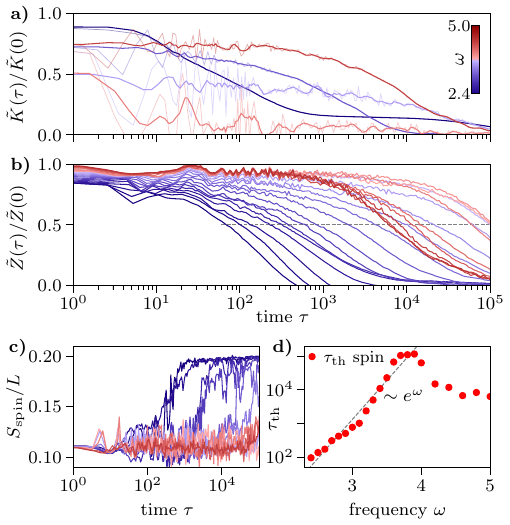}
\caption{\textbf{Fractionalized prethermalization with exponential lifetime:}  (a) While heating occurs rather slowly in the charge sector at small frequencies, we see an opposite trend with increasing frequency: Heating rates in the charge sector increase up to frequencies $\omega \approx 4t$ as captured by the kinetic energy. For better visibility, we show data for a few frequencies and apply a moving average conserving filter (unfiltered data in light colors below). (b) The spin correlations show that heating is strongly suppressed in the spin sector leading to a fractionalized prethermal plateau with exponential lifetime at smaller frequencies, before approaching a constant value determined by the quasiparticle lifetime. (c) The prethermal nature of the spin state is also captured by the half-chain entanglement entropy of the squeezed spin chain. (d) At intermediate frequencies, the lifetime of the fractional prethermal plateau shows exponential dependence on frequency $\omega$. Numerical data is obtained for $U=26t$.}
\label{fig:staggered-hopping-drive}
\end{figure}

The second scenario we consider is a drive that couples asymmetrically to the chargon and spinon. As an example, we drive the underlying Hubbard model with an oscillatory staggered hopping term. The action of this drive on the low energy effective theory is derived in detail in Sec.$\,$\ref{sec:Floquet-expansion-Hubbard} by introducing a time-dependent variant of the Schrieffer-Wolff transformation. This leads to an unusual time-dependent  effective Hamiltonian that  in the low-energy subspace reads
\begin{eqnarray}
    \begin{aligned}    
    &gf(\tau)\hat{V} = \sum_j J'_j(\tau) \bigg( \mathbf{S}_{j}\mathbf{S}_{j+1} - \frac{n_{j}n_{j+1}}{4} \bigg) \\
    &+gf(\tau)\mathcal{P}_0 \bigg[ \sum_{j,\alpha=\uparrow, \downarrow}(-1)^j c^\dagger_{j,\alpha}c_{j+1, \alpha} + \mathrm{H.c.}) \bigg] \mathcal{P}_0,
    \end{aligned}
    \label{eq:staggered-hopping-drive}
\end{eqnarray}
where 
\begin{eqnarray}
    J'_j(\tau) = -\frac{2gJ}{t}f(\tau)(-1)^j+\frac{g^2J}{t^2}f(\tau)^2.
\end{eqnarray}
As before, $f(\tau)$ is chosen to be piecewise constant in numerical simulations. The representation in squeezed space is given in Appendix \ref{A:squeezed-space}. Crucially, at moderate drive strength $g\sim t$, the drive couples to the chargon with strength $\mathcal{O}(t)$ and to the spinon with strength $\mathcal{O}(J)$. This can be intuitively understood by noting that the only drive term of order $\mathcal{O}(t)$ is a staggered hopping, which corresponds to charge defects hopping on the links of the spin chain. Consequently, the drive couples to both quasiparticles at moderate amplitude compared to their intrinsic energy scale, but with different relative frequency. At moderate frequencies, we again expect fast heating in the charge sector due to the drive coupling (strongly) to the chargon via the staggered hopping term. Contrarily, the spin sector is driven at much higher relative frequencies compared to the spin exchange energy $J$, which can result in a "conventional" prethermal plateau \cite{bukov2015universal} within the sector with exponential lifetime in frequency \cite{kuwahara2016floquet}, again leading to a fractionalized prethermal plateau; see \rfig{\ref{fig:staggered-hopping-drive}}(a,b).

\begin{figure}
\includegraphics{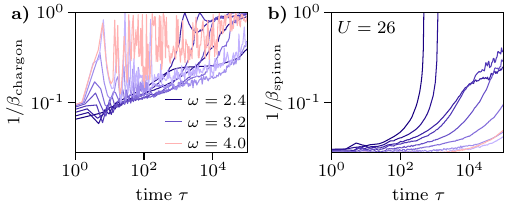}
\caption{\textbf{Heating dynamics:} At small frequencies, heating in the charge and spin sector show opposite trends: a) Energy absorption in the charge sector increases with frequency, which is captured by an increasing effective temperature of the kinetic energy $\tilde{K}(\tau)$. b) By contrast, as expected by prethermalization arguments, energy absorption in the spin sector is greatly reduced, as captured by the low effective temperature of the spins. Numerical data is obtained for $U=26t$ and initial temperature $1/\beta_0 = 0.08t$. Both charge and spin temperatures are vastly different.}
\label{fig:staggered-hopping-drive-temperatures}
\end{figure}

So far, we have ignored the finite quasiparticle lifetime. Due to the exponential lifetime $\tau_\text{pre}\sim e^{\mathcal{O}(\omega/J)}$ of the prethermal plateau compared to the quasiparticle lifetime $t\tau_0 \sim t^2/J^2$, we generally expect that the lifetime of the fractionalized prethermal plateau is cut off by the latter. In particular, we expect that the exponential scaling is not visible in the limit $U\rightarrow \infty$. However, as shown in \rfig{\ref{fig:staggered-hopping-drive}}(d), for intermediate $U$, the exponential scaling is still clearly visible in a frequency range $\omega \in (2.4,4.0)$. The dynamic frequency range where the exponential scaling is visible is comparably small, which is mainly caused by the charge sector starting to prethermalize for frequencies $\omega \geq 4.2$. Additionally, the integrable charge sector may heat to a generalized Gibbs ensemble rather than a true infinite temperature state  for small frequencies. However, one can clearly see an opposing trend between the heating of the charge and spin sector. While the former absorbs energy more efficiently with increasing frequency, heating in the latter is greatly suppressed. We show this heating trend in \rfig{\ref{fig:staggered-hopping-drive-temperatures}}. Additionally, for the largest frequencies considered, we observe that the prethermal plateau in the spin sector approaches a constant lifetime, which is comparable to the lifetime found in Section \ref{sec:frac-pretherm-potential}. Thus, we confirm the expectation that the static coupling of quasiparticles serves as a cutoff to the lifetime of the fractionalized prethermal plateau.

The prethermal character of the spin state can be further established by investigating the behavior of the half chain entanglement entropy of the squeezed spin chain
\begin{eqnarray}
\begin{aligned}
    S_\text{spin}(\tau) &= -\text{Tr}  (\rho_B(\tau)\log(\rho_B(\tau))) \\
    \rho_B(\tau) &= \text{Tr}_{1,...,N/2}(\rho_s(\tau)) \\
    \rho_s(\tau) &= \text{Tr}_\text{holes}(|\psi(\tau)\rangle\langle \psi (\tau) |). 
\end{aligned}
\end{eqnarray}
The half-chain spin entropy approaches its maximal value corresponding to a thermal (or random) state at times comparable to the heating times of the spin sector, see \rfig{\ref{fig:staggered-hopping-drive}}(c). By contrast, throughout the transient plateau, it remains close to its initial value, confirming the prethermal character of the plateau.

We note that the half chain entropy of the \textit{full} system shows a stair-case like behavior as in Ref. \cite{jin2023fractionalized}, thus also capturing the heating of the charge sector. In Appendix \ref{A:additional-data} we additionally numerically confirm Class (III), that the fractional prethermal plateau vanishes when the drive couples strongly (i.e., with strength $\mathcal{O}(t)$) to both sectors.

\section{Floquet expansion of the Hubbard model}
\label{sec:Floquet-expansion-Hubbard}

In the following we argue that the drives used in Sec.$\,$\ref{sec:fractionalized-prethermalization} correspond to different classes of drives that arise in the effective low-energy description. We are interested in the range of intermediate frequencies, where the effective Floquet Hamiltonian is highly nonlocal and the usual high frequency expansions \cite{bukov2015universal,kuwahara2016floquet, bukov2016schrieffer} break down. Instead, we propose a scheme based on a time-dependent variant of the Schrieffer-Wolff (SW) transformation to determine the low-energy effective theory. Our approach is loosely related to Ref. \cite{deRoeck2019very}. For a more comprehensive guide to the method, we first briefly recall the standard SW transformation \cite{bravyi2011schrieffer} as a way to derive the tJ-model in \req{\ref{eq:tJ-model}} as the low-energy effective theory of the Hubbard model and only afterwards generalize the method to treat driven Hubbard models.

In the most general case the SW rotation is applied to a Hamiltonian of the form
\begin{eqnarray}
    H = H_0 + \epsilon \mathcal{V}
    \label{eq:H-general-SW-timeless}
\end{eqnarray}
that consists of a "simple" part $H_0$ that is diagonal in some basis $|n\rangle$ and a small perturbation $\epsilon \mathcal{V}$. Additionally, the eigenvalues of $H_0$ are required to decompose into blocks separated by a large energy scale $U$. The goal is to find a transformation $e^S$ such that the transformed Hamiltonian
\begin{eqnarray}
    H_\mathrm{eff} = e^SHe^{-S}
    \label{eq:time-independent-SW}
\end{eqnarray}
is block diagonal in the rotated basis $|n'\rangle=e^S|n\rangle$ up to some order in $\epsilon$. The generator $S$ may be chosen purely offdiagonal and antihermitian, and for large enough $U$, it can be shown \cite{bravyi2011schrieffer} that the series expansion $S=\sum_n \epsilon^nS_n$ in powers of $\epsilon$ is well defined. Subsequently, one can expand \req{\ref{eq:time-independent-SW}} using the Baker-Campbell-Hausdorff formula
\begin{eqnarray}
   H_\text{eff} = e^SHe^{-S} = H + [S,H]+\frac{1}{2}[S,[S,H]]+...
   \label{eq:time-independent-SW-BCH-expansion}
\end{eqnarray}
and order the resulting equation in powers of $\epsilon$. The components $S_n$ can now be determined recursively by requiring that the block-offdiagonal contributions vanish. The remaining diagonal terms constitute the effective Hamiltonian up to order $\mathcal{O}(\epsilon^n)$.

As an example, let us briefly revisit the strong coupling limit of the Hubbard model from \req{\ref{eq:Hubbard-model}}, where the diagonal part $H_0$ is given by the onsite interaction and the hopping is treated as perturbation
\begin{eqnarray}
    \begin{aligned}
        H_0 &= \frac{U}{2}\sum_j n_j(n_j-1) \\
        \epsilon\mathcal{V} &= -t\sum_{\langle i,j\rangle,\alpha} c^\dagger_{i,\alpha}c_{j+1,\alpha}.
    \end{aligned}
\end{eqnarray}
The point of interest is the effective Hamiltonian in the lowest lying subspace of $H_0$ which is composed of states with at most one particle per site, such that $H_0 = 0$. The terms to linear order in $\epsilon$ appearing in \req{\ref{eq:time-independent-SW-BCH-expansion}} are
\begin{eqnarray}
    \begin{aligned}
    \mathcal{V}+[S_1,H] = &\mathcal{P}_0\mathcal{V}\mathcal{P}_0+\mathcal{P}_c\mathcal{V}\mathcal{P}_c +  \\
    &\mathcal{P}_0\mathcal{V}\mathcal{P}_c+\mathcal{P}_c\mathcal{V}\mathcal{P}_0+[S_1,H]
    \end{aligned}
    \label{eq:standard-SW-1st-offdiagonal}
\end{eqnarray}
where we have introduced the projector $\mathcal{P}_0$ that projects onto the space with at most one particle per site and its complement $\mathcal{P}_c$. Next $S_1$ has to be chosen such that the offdiagonal hopping terms which couple sectors with a different number of double occupied sites
\begin{eqnarray}
    0 = \mathcal{P}_0\mathcal{V}\mathcal{P}_c+\mathcal{P}_c\mathcal{V}\mathcal{P}_0+[S_1,H_0]
    \label{eq:1st-order-SW-condition}
\end{eqnarray}
are eliminated. Employing the general formalism introduced in \cite{bravyi2011schrieffer} shows that this can be achieved by choosing
\begin{eqnarray}
    S_1 = \frac{1}{U}(\mathcal{P}_c\mathcal{V}\mathcal{P}_0-\mathcal{P}_0\mathcal{V}\mathcal{P}_c).
    \label{eq:S1-Hubbard-model}
\end{eqnarray}
Subsequently, the same procedure is to be repeated with the terms quadratic in $\epsilon$. To recover the tJ-model, it is sufficient to consider the block-diagonal terms in second order 
\begin{eqnarray}
    \begin{aligned}
        & [S_1,\mathcal{P}_0 \mathcal{V} \mathcal{P}_c+\mathcal{P}_c \mathcal{V} \mathcal{P}_0]+\frac{1}{2}[S_1,[S_1,H_0]] =  \\
    & - \frac{1}{U} ( \mathcal{P}_0 \mathcal{V} \mathcal{P}_c \mathcal{V} \mathcal{P}_0 - \mathcal{P}_c \mathcal{V} \mathcal{P}_0 \mathcal{V} \mathcal{P}_c).
    \end{aligned}
    \label{eq:second-order-contribution-base}
\end{eqnarray}
Together with the remaining first order contributions, one recovers  
\begin{eqnarray}
    H_\mathrm{eff} = -t \mathcal{P}_0 \mathcal{V} \mathcal{P}_0 - \frac{t^2}{U} \mathcal{P}_0 \mathcal{V} \mathcal{P}_c \mathcal{V} \mathcal{P}_0
\end{eqnarray}
as the effective Hamiltonian in the lowest sector to second order in $t$. Here, we can recognize the hopping term in the first term. The second term accounts for second order exchange processes of the form 
\begin{eqnarray}
    |\bar{\alpha},\alpha\rangle \rightarrow |2,0\rangle \rightarrow 
    \begin{cases}
        |\bar{\alpha}, \alpha \rangle \\
        -|\alpha, \bar{\alpha} \rangle
    \end{cases}
    \; \; \; \alpha = \uparrow, \downarrow
\end{eqnarray}
which upon appropriate rearrangement result in the effective spin-spin interaction from \req{\ref{eq:tJ-model}}. Here, we again neglect "hop-over" processes of the form $|0,\bar{\alpha},\alpha\rangle \rightarrow |0,2,0\rangle \rightarrow |\alpha,\bar{\alpha},0\rangle$ which constitute a subleading correction to the tJ-model and are suppressed in the hole doped regime.

We now generalize the formalism to driven Hubbard models. We require that the drive amplitude is in the order of the hopping strength $t$ such that the time-dependent part can be absorbed into the perturbation. The driven Hubbard Hamiltonian can then be written as 
\begin{eqnarray}
    H(\tau) = \frac{U}{2}\sum_i n_i(n_i-1) + \epsilon \mathcal{V}(\tau) \equiv H_0 + \epsilon \mathcal{V}(\tau)
    \label{eq:driven-Hubbard-SW}
\end{eqnarray}
where $\mathcal{V}(\tau)$ accounts for the hopping and applied drive. Again, the aim is to bring the Hamiltonian into a block diagonal form with respect to the onsite interaction, where now we make use of a time dependent transformation of the form $|\psi'(\tau)\rangle=e^{S(\tau)}|\psi(\tau)\rangle$. We emphasize that $S(\tau)$ does not necessarily commute with itself at different times, but may be chosen purely offdiagonal and antihermitian as in standard SW transformations. The transformed state then fulfills the Schrödinger equation
\begin{eqnarray}
    i\partial_\tau |\psi'(\tau)\rangle = H_\text{eff} |\psi'(\tau)\rangle
\end{eqnarray}
with 
\begin{eqnarray}
    H_\text{eff} = i (\partial_\tau e^{S(\tau)})e^{-S(\tau)}+e^{S(\tau)}He^{-S(\tau)}.
    \label{eq:Heff-time-dependent}
\end{eqnarray}
As in the case of a time independent rotation, we would like to expand $S(\tau) = \sum_n \epsilon^n S_n(\tau)$ in powers of epsilon and compute the contributions recursively. The second contribution to $H_\text{eff}$ in \req{\ref{eq:Heff-time-dependent}} can again be expanded using the Baker-Campbell-Hausdorff formula. In order to treat the integral, we first use the identity (\cite{wilcox1967exponential}, Eq.$\,$(2.1))
\begin{eqnarray}
    i(\partial_\tau e^{S(\tau)})e^{-S(\tau)} = i \int_0^1e^{\lambda S(\tau)}\partial_\tau S(\tau) e^{-\lambda S(\tau)} d\lambda,
\end{eqnarray}
to represent the derivative in integro-differential form. We can then expand both exponentials in the integral to carry out integration and order the resulting series in powers of $\epsilon$ by inserting the series expansion for $S(\tau)$
\begin{eqnarray*}
    \begin{aligned}
        i &\int_0^1e^{\lambda S(\tau)}\partial_\tau S(\tau) e^{-\lambda S(\tau)} d\lambda = \\
    i &\int_0^1 \partial_\tau S(\tau)+\lambda \big [ S(\tau)\partial_\tau S(\tau)-(\partial_\tau S(\tau))S(\tau) \big] + ... \;  d\lambda = \\
    &i\epsilon \partial_\tau S_1(\tau)+i\epsilon^2 \big [ \partial_\tau S_2(\tau) + \frac{1}{2}S_1(\tau)\partial_\tau S_1(\tau) - \\&\frac{1}{2}(\partial_\tau S_1(\tau))S_1(\tau) \big ] + \mathcal{O}(\epsilon^3).
    \end{aligned}
\end{eqnarray*}
In similar style to the standard SW transformation, the components $S_n(\tau)$ can then be determined iteratively in powers of $\epsilon$ by requiring that the offdiagonal terms vanish. The resulting equations to determine $S_n(\tau)$ have the general structure
\begin{eqnarray}
    i\partial_\tau S_n(\tau) + [S_n(\tau), H_0] + \mathcal{R}_n(\tau) = 0
    \label{eq:time-dependent-SW-general-condition}
\end{eqnarray}
where $\mathcal{R}_n(\tau)$ does not depend on $S_n(\tau)$ and includes all offdiagonal terms of order $\mathcal{O}(\epsilon^n)$ created by the nested commutators of $S_m(\tau),$ $m<n$ with $H(\tau)$ as well as offdiagonal terms resulting from the derivative. Instead of a set of algebraic equations, one is left with a first order inhomogeneous linear differential equation with time-dependent coefficients. Existence of a unique solution for this type of equation is guaranteed by the theorem of Picard-Lindelöf under reasonable physical assumptions on the matrix coefficients, justifying the recursive scheme. Nevertheless, these equations are generally hard to solve in practice.

In the following, we argue that the above introduced scheme can still be useful to treat a variety of systems. Let us start by giving some remarks: First, we note that the drive in \req{\ref{eq:driven-Hubbard-SW}} does not generate additional offdiagonal terms if it is diagonal in the basis of the onsite interaction $H_0$. Consequently, the drive does not induce higher-order exchange processes and the rotation $S(\tau)$ may be chosen time independent. This is for example the case for the staggered potential used in Sec.$\,$\ref{sec:frac-pretherm-potential} as well as the nearest-neighbor spin-z interaction used  in Appendix \ref{A:additional-data}. 

Secondly, let us take a closer look at the equation determining $S_1(\tau)$
\begin{eqnarray}
    i\partial_\tau S_1(\tau) + \mathcal{V}(\tau) + [S_1(\tau),H_0] = 0.
    \label{eq:time-dependent-SW-1st-order}
\end{eqnarray}
The structure of the equation determining $S_1(\tau)$ differs slightly, since there are no residual terms $\mathcal{R}_1(\tau)=0$ from lower order contributions, but the perturbation $\mathcal{V}(\tau)$ originating from the Hamiltonian $\mathcal{H}(\tau)$ has to be included. The term $H_0$ is of order $\mathcal{O}(U)$, whereas $\mathcal{V}(\tau)$ is of order $\mathcal{O}(t)$. Ignoring the derivative for the moment, this implies that $S_1(\tau)$ is of order $\mathcal{O}(t /U)$, implying that the derivative is smaller by order $\mathcal{O}(\omega \cdot  t/U)$ compared to the other terms. In the relevant parameter regime $\omega \sim t$ and $U \gg  t$, we therefore expect the corrections to $S_1(\tau)$ caused by the derivative to be of order $\mathcal{O}(\omega \cdot t /U^2)$. Consequently, using \req{\ref{eq:second-order-contribution-base}}, the largest relative corrections to the second order exchange processes are of order $\mathcal{O}(\omega/U)$. The analogue consideration can be made for the components $S_n(\tau)$: Given the residual in \req{\ref{eq:time-dependent-SW-general-condition}} is of order $\mathcal{O}(r)$, $S_n(\tau)$ has to be of order $\mathcal{O}(r/(\omega+U))$ with the leading relative correction caused by the derivative being $\mathcal{O}(\omega/U)$ again. The key consequence is that corrections caused by the time derivative are generally small. Thus, ignoring the derivative is a reasonable approximation in the limit $U\to \infty$, which is equal to assuming that the generator $S(\tau)$ commutes with itself at different times. Consequently, \req{\ref{eq:time-dependent-SW-general-condition}} are reduced to a set of algebraic equations.

As an example, we consider driving the Hubbard model with a staggered hopping of the form
\begin{eqnarray}
    V = \sum_{j,\alpha=\uparrow, \downarrow}(-1)^j(c^\dagger_{j,\alpha}c_{j+1, \alpha} + \mathrm{H.c.})
\end{eqnarray}
to show how the above considerations appear in practice. In this case
\begin{eqnarray}
    \epsilon \mathcal{V}(\tau) = -t\sum_{j,\alpha=\uparrow, \downarrow}A_{j,\alpha}(\tau)(c^\dagger_{j,\alpha}c_{j+1, \alpha} + \mathrm{H.c.})
\end{eqnarray}
where $A_{j,\alpha}(\tau) = 1-\frac{g}{t}f(\tau)(-1)^j$.
We compute the effective Hamiltonian in the lowest sector using \req{\ref{eq:Heff-time-dependent}} and \req{\ref{eq:time-dependent-SW-general-condition}}. Here, we resort to stating the main results, the calculations are given in more detail in Appendix \ref{A:Schrieffer-Wolff}. First, we require that the offdiagonal terms to first order in $\epsilon$ resulting from \req{\ref{eq:time-dependent-SW-1st-order}} vanish, which leads to the condition
\begin{eqnarray}
    \begin{aligned}
    &i\partial_\tau S_1(\tau)+ [S_1(\tau),H_0] + \\
    &\mathcal{P}_0 \mathcal{V}(\tau) \mathcal{P}_c+\mathcal{P}_c \mathcal{V}(\tau) \mathcal{P}_0 = 0
    \end{aligned}
    \label{eq:SW-time-dependent-1st-order-OD}
\end{eqnarray}
where we use the same convention for the projectors as in \req{\ref{eq:standard-SW-1st-offdiagonal}}. 

If we ignore the derivative, \req{\ref{eq:SW-time-dependent-1st-order-OD}} is solved by 
\begin{eqnarray}
    S_1(\tau) = \frac{1}{U}[\mathcal{P}_c\mathcal{V}(\tau)\mathcal{P}_0-\mathcal{P}_0\mathcal{V}(\tau)\mathcal{P}_c]
    \label{eq:S1-driven-no-dt}
\end{eqnarray}
and is thus of the same stucture as \req{\ref{eq:S1-Hubbard-model}}.
Next, we include the derivative in our calculations. For general $f(\tau)$, finding an analytic solution for \req{\ref{eq:SW-time-dependent-1st-order-OD}} is not possible. The equation can however be solved for $f(\tau)=\sin(\omega\tau)$, which we use as an example in the following. The resulting $S_1(\tau)$ has the same structure as \req{\ref{eq:S1-driven-no-dt}}, but with slightly modified coefficients
\begin{eqnarray*}
        B_{j,\alpha}(\tau )=1 - (-1)^j \Big( \frac{gU^2\sin(\omega\tau)}{t(U^2-\omega^2)}+\frac{igU\omega\cos(\omega\tau)}{t(U^2-\omega^2)} \Big)
        \label{eq:coefficients-S1-with-dt}
\end{eqnarray*}
instead of $A_{j,\alpha}(\tau)$. From the above equation, the relative corrections of order $\mathcal{O}(\omega/U)$ are immediately apparent. First, the prefactor of the sine part changes from $g/t  \rightarrow \frac{g}{t(1-\omega^2/U^2)} \approx \frac{g}{t}(1+\omega^2/U^2)$ and additionally, the coefficients acquire a small cosine contribution of order $\mathcal{O}(\omega/U)$.

Subsequently, the effective Hamiltonian to second order has to be determined from \req{\ref{eq:Heff-time-dependent}}. For the case where the derivative was ignored and using \req{\ref{eq:S1-driven-no-dt}}, we recover as the effective Hamiltonian up to second order
\begin{widetext}
    \begin{eqnarray}
        H(\tau) = H_{tJ} + gf(\tau)\mathcal{P}_0 \bigg[ \sum_{j,\alpha=\uparrow, \downarrow}(-1)^j c^\dagger_{j,\alpha}c_{j+1, \alpha} + \mathrm{H.c.}) \bigg] \mathcal{P}_0 + \sum_j J'_j(\tau) \bigg( \mathbf{S}_{j}\mathbf{S}_{j+1} - \frac{n_{j}n_{j+1}}{4} \bigg).
    \end{eqnarray}
\end{widetext}
where 
\begin{eqnarray}
    J'_j(\tau) = -\frac{2gJ}{t}f(\tau)(-1)^j+\frac{g^2J}{t^2}f^2(\tau),
    \label{eq:Jprime-base}
\end{eqnarray}
and we recover \req{\ref{eq:staggered-hopping-drive}} up to normalization of $J \rightarrow \frac{4(t^2+g^2)}{U}$. 
Reintroducing the derivative, the parameter $J'_j(\tau)$ from \req{\ref{eq:Jprime-base}} is slightly modified to
\begin{eqnarray}
    \frac{1+J'_j(\tau)}{J} = 2\text{Re}(B_{j,\alpha})A_{j,\alpha}-|B_{j,\alpha}|^2.   
\end{eqnarray}
The calculation of the resulting terms is straightforward and the precise form is given in Appendix \ref{A:Schrieffer-Wolff}. Here, we merely emphasize that corrections to the second order exchange strength are of order $\mathcal{O}(\omega^2/U^3)$.

To summarize, the explicit time dependence of the drive induces corrections to the effective Hamiltonian in the lowest sector. Since these corrections are  generally of order $\mathcal{O}(\omega/U)$ relative to the unperturbed term, they vanish in the limit $U\to \infty$ and are thus minor in the regime $U\gg \omega, t$. The above formalism also allows for a classification of drives regarding the existence and lifetime of a fractional prethermal plateau as shown in \rfig{\ref{fig:basics-sketch1}}(b). Generally, a fractional prethermal plateau can only occur if the periodic modulation couples to the spinon only via higher order exchange processes. A particular case are drives that are diagonal in the basis of $H_0$ as outlined before, as they do not couple to the spinon via higher order processes at all and lead to a fractional prethermal plateau with lifetime determined by the quasiparticle lifetime largely independent of frequency. 

As a last remark, let us mention that \req{\ref{eq:Heff-time-dependent}} can in principle be solved to arbitrary order in $\epsilon$ similar to the standard SW transformation.  Prethermalization arguments then suggest that the resulting effective low energy theory is valid up to exponential times: For the standard SW transformation, the exponential lifetime can be nicely captured via a transformation into the rotating frame and a subsequent high frequency expansion that yields the same result \cite{bukov2015universal}. For the time-dependent SW transformation, the exponential validity was formally derived in \cite{deRoeck2019very}. We also note an analogy to multiband systems, where prethermalization is strongly suppressed when the driving frequency is comparable to the bandgap \cite{sun2020optimal} due to interband processes becoming relevant. However, in contrast to multiband systems, fractionalized prethermalization emerges in a system with a single physical degree (the electron), which due to the interacting nature of the problem fractionalizes into spinon and holon. Thus even though physical degrees of freedom are driven, a highly complex state is established, which cannot be captured by a simple effective Hamiltonian.

\section{Experimental Prospects}

The Hubbard model from \req{\ref{eq:Hubbard-model}} is naturally realized by loading an optical lattice with ultracold atoms \cite{bloch2008many}. The lattice depth and lattice constant can be used to control the hopping strength of atoms between lattice sites, thus providing high tunability of the ratio $U/t$ of onsite interaction and hopping amplitude.

The periodic modulations from \req{\ref{eq:staggered-potential-drive}} and \req{\ref{eq:staggered-hopping-drive}} can be realized by adiabatic modulation of an additional laser potential $V_d$ of wavelength $2\lambda$:
\begin{eqnarray}
    \begin{aligned}
    V(x) &= E_r (s_1V_0 + s_2(\tau)V_d) \\
    V_0 &= \sin^2(k_Lx) \\
    V_{d} &= \sin^2(\frac{k_L}{2}x+\phi_d)
    \end{aligned}
    \label{eq:lattice-potential}
\end{eqnarray}
where $k_L=2\pi/\lambda$ is the lattice wave vector, $E_r = \frac{\hbar^2k_L^2}{2m}$ the recoil energy, $m$ the mass of the atoms and $s_2(\tau) = s_2\sin(\omega\tau)$ the time dependent superlattice amplitude. From this, we evaluate the modulation of the hopping paramters and the interaction strength by solving the band structure numerically, see Appendix \ref{A:single particle SE}. 

For $\phi_d=0$ and small $s_2 \lesssim  0.2$, \req{\ref{eq:lattice-potential}} realizes the staggered potential term in \req{\ref{eq:staggered-potential-drive}}. The superlattice introduces and energy offset $\delta V = s_2$ between even and odd lattice sites, while maintaining approximately constant lattice depth and thus nearly constant hopping and onsite interaction strength.

To realize a strong staggered hopping as in \req{\ref{eq:staggered-hopping-drive}}, we set $\phi_d=\pi/4$ and $s_2 \approx s_1/2$. This generates an array of double wells with alternating barrier heights. In this setup, every second hopping element is suppressed due to the exponential dependence on lattice depth, giving rise to a staggered hopping configuration. Modulating the lattice depth also affects hopping and onsite interactions. To counteract this effect, we stabilize the average parameters by dynamically adjusting the primary lattice depth as $s_1(\tau) = s_1+s_2(\tau)/2.7$. This simple protocol ensures that the average hopping and onsite interaction strengths remain stable within a 5\% deviation, see Appendix \ref{A:single particle SE}. 

In order to capture fractionalized prethermalization experimentally, a low temperature initial state needs to be prepared. Subsequently, the system is periodically driven with one of the potentials introduced above. The fractional prethermal plateau is then captured by evaluating the spin correlations and kinetic energy at different points in time. The former can be measured via single site resolved fluorescence imaging of the spin particles \cite{wei2022quantum}. The latter can also be measured in cold atom experiments \cite{wybo2023preparing, impertro2023local} albeit through a slightly more elaborate scheme. The general idea is to introduce large potential barriers after every second site, effectively confining the particles to a series of double wells. The double well can subsequently be interpreted as a two level system by introducing an offset between the minima and the kinetic energy over the bond can be evaluated via readout of the x-component of the two level system.

\begin{figure}
\includegraphics{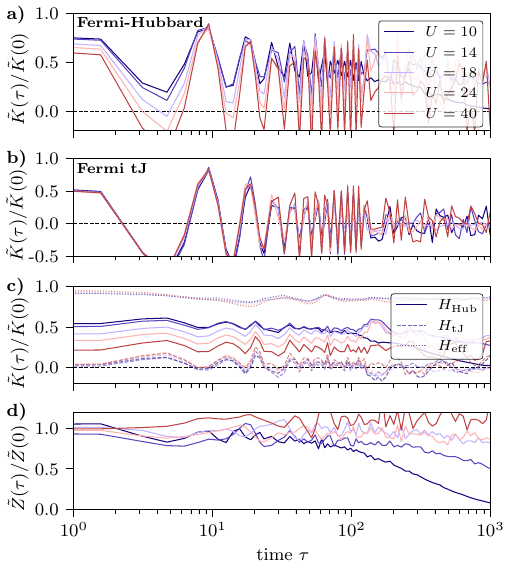}
\caption{\textbf{Fractionalized prethermalization in a driven Hubbard model:} The dynamics of the kinetic energy (a) of a driven Hubbard model qualitatively capture the respective physics of the low energy effective tJ-model (b). In (c) we show that the resulting dynamics cannot be described by an effective Hamiltonian obtained from the high frequency Magnus expansion and the dynamics of the Hubbard and tJ-model approaches with increasing $U$. To improve clarity, we apply a moving average preserving filter to smoothen the data in (c). Additionally, the spin correlations of a driven Hubbard model retains most of its initial value over an extended period of time. Combining these observations demonstrates fractional prethermalizaiton in the Hubbard model.}
\label{fig:staggered-pot-Hubbard}
\end{figure}

A natural question is to which extent the effective description of the tJ-model can be observed in cold atom experiments implementing Hubbard Hamiltonians. Determining the lifetime of prethermal plateaus in general is an experimental challenge \cite{rubioAbadal2020floquet}. In \rfig{\ref{fig:staggered-pot-Hubbard}}, we show that the key features of a fractional prethermal plateau are  qualitatively captured in the dynamics of a driven Hubbard model as well. Specifically, we consider the driven staggered potential term of \req{\ref{eq:staggered-potential-drive}}, giving rise to the prethermalization scenario Case (I). Due to the larger local Hilbert space, we are restricted to small systems of size $L=14$ with $N=12$ particles. Nevertheless, we find that the dynamics of the kinetic energy generated by the Hubbard model qualitatively match the dynamics of the respective effective tJ-model. In particular, agreement improves with increasing onsite interaction $U$, as expected. Additionally, the spin correlations retains most of its initial value and the relaxation time becomes significantly larger upon increasing $U$. For comparison, we also include the dynamics generated by the corresponding second order Magnus expansion \cite{bukov2015universal} to show that the Hubbard model physics are not correctly captured by this high-frequency expansion, implying that the resulting state does not have conventional prethermal character.

\section{Summary \& Outlook}

We have explored fractional prethermalization in one-dimensional tJ- and Hubbard-models, originating from spin-charge separation. We argue that the respective physics in the two sectors can be captured by the kinetic energy and spin-spin couplings respectively. Resorting to different types of the drive, we showed that fractional prethermal plateaus with both exponential lifetime in frequency as well as Fermi-Golden rule lifetime in the onsite interaction strength can be realized. Independent of the specific drive, the fractional prethermal plateau is characterized by a high effective temperature in the charge sector, coexisting with a low temperature prethermal state in the spin sector. We further showed that such physics arise in the low-energy effective theory of one dimensional hole doped Hubbard models by employing a time dependent variant of the Schrieffer-Wolff transformation. Similar results are also expected for bosonic Hubbard and tJ-models. The introduced methods allows us to classify drives according to the existence and lifetime of the fractional prethermal plateau. Drives that couple to the spinon only via higher order exchange processes generally lead to a fractional prethermal plateau, while drives that couple strongly to both quasiparticles generally cause a breakdown of spin-charge separation, and consequently do not result in a fractional prethermal plateau. 

Our work demonstrates that strongly correlated systems with fractionalized excitations can host intriguing prethermal Floquet states beyond the conventional paradigms and that such states offer a promising tool to probe fractionalization in such systems. For future works it would be interesting to identify other systems showing unconventional prethermalization behavior beyond the one-dimensional Hubbard model and the Kitaev spin liquid \cite{jin2023fractionalized} as well as exploring the interplay between fractionalization and other non-equilibrium phenomena such as time crystals \cite{else2020discrete,zhang2017observation,zaletel2023colloquium} and topological Floquet systems \cite{will2025probing}. Furthermore, another exciting direction is to explore how in quantum information processing architectures Trotterization protocols to study
the dynamics of excitations in (1+1)D \cite{martinez2016real} and (2+1)D \cite{cochran2024visualizing} lattice gauge theories could lead to related fractionalized prethermalization phenomena.

\begin{acknowledgments}
We thank Marin Bukov, David Wei, Johannes Zeiher and Philip Zechmann for stimulating discussions. Simulations within this work are based on on the quspin package \cite{weinberg2019quspin}. We acknowledge support from the Deutsche Forschungsgemeinschaft (DFG, German Research Foundation) under Germany’s Excellence Strategy–EXC–2111–390814868, TRR 360 – 492547816 and DFG grants No. KN1254/1-2, KN1254/2-1, the European Research Council (ERC) under the European Union’s Horizon
2020 research and innovation programme (grant agreement No 851161), the European Union (grant agreement No 101169765), as well as the Munich Quantum Valley, which is supported by the Bavarian state government with funds from the Hightech Agenda Bayern Plus. 
\end{acknowledgments}

\section*{Data avaliability}
Numerical data and codes that support the findings of this article are available on Zenodo and from the authors upon reasonable request \cite{zenodo}.

\appendix

\section{Squeezed space formalism}
\label{A:squeezed-space}
Here, we give a more detailed overview on the squeezed-space formalism following Ref. \cite{bohrdt2018angle} to analyze the spin-charge separation in the tJ model \req{\ref{eq:tJ-model}}.

We first consider a single hole and introduce chargons $h_j$ and spinons $f_{j,\alpha}$ to describe the hole and spin degrees of freedom such that
\begin{eqnarray}
    c_{j,\alpha} &=& h_j^\dagger f_{j,\alpha} \\
    \mathbf{S}_j &=& \frac{1}{2} \sum_{\alpha,\beta} f^\dagger_{i,\alpha} \mathbf{\sigma}_{\alpha,\beta}f_{j,\beta}
\end{eqnarray}
with the condition
\begin{eqnarray}
    \sum_\alpha f^\dagger_{j,\alpha}f_{j,\alpha}+h_j^\dagger h_j = 1.
    \label{eq:constraint}
\end{eqnarray}
In terms of the new operators the original hopping term acquires a quartic form
\begin{eqnarray}
    \mathcal{T} = -t\sum_{j,\alpha} (f_{j,\alpha}^\dagger h_j h_{j+1}^\dagger  f_{j+1,\alpha} + \mathrm{H.c.})
\end{eqnarray}
Further simplification is subsequently achieved by considering only spinons on the occupied sites $\tilde{j} = j + \sum_{i \leq j} h_i^\dagger h_i$, leading to new operators $\tilde{f}_{j,\alpha}=f_{\tilde{j},\alpha}$. The transformation is equivalent to mapping the chargon onto the bonds, eliminating it from the constraint Eq.~\eqref{eq:constraint} which reduces to
\begin{eqnarray}
    \sum_\alpha \tilde{f}^\dagger_{j,\alpha}\tilde{f}_{j,\alpha} = 1.
\end{eqnarray}
The new basis states are now states of the form $h^\dagger_j|0\rangle \otimes \prod_{j}\tilde{f}^\dagger_{j,\alpha_{j}}|0,0,0,...,0\rangle$, i.e. the Hilbert space is described by two distinguishable particles with $[\tilde{f},h]=0$. In terms of the new operators one has
\begin{eqnarray}
    \mathcal{T} &=& -t\sum_{j,\alpha} (f_{j,\alpha}^\dagger h_j h_{j+1}^\dagger f_{j+1,\alpha} + \mathrm{H.c.} )\nonumber \\
    &=& -t \sum_{j,\alpha} (\tilde{f}_{j,\alpha}^\dagger h_{j} h_{j+1}^\dagger \tilde{f}_{j, \alpha} + \mathrm{H.c.}) \nonumber \\
    &=& -t \sum_{j}  (h_{j} h_{j+1}^\dagger \underbrace{\sum_\alpha \tilde{f}_{j,\alpha}^\dagger \tilde{f}_{j, \alpha}}_{=1} + \mathrm{H.c.})
    \label{eq:kinetic-energy-squeezed}
\end{eqnarray}
where in the second line it is used that the chargon resides on site $j+1$ upon application of $f^\dagger_{j,\alpha}$.

Additionally, the spin dependent part is conveniently expressed as
\begin{eqnarray}
\mathcal{S} = J \sum_j \tilde{\mathbf{S}}_{j}\tilde{\mathbf{S}}_{j+1}\cdot(1-h_j^\dagger h_j)
\label{eq:spins-squeezed}
\end{eqnarray}
since the two spinons adjacent to the chargon are not neighboring spins in the original basis and therefore do not interact. The sum of density-density interactions is constant. Combining these terms recovers the transformed Hamiltonian of Ref. \cite{bohrdt2018angle} 
\begin{eqnarray*}
    H = -t\sum_j (h^\dagger_j h_{j+1} +\mathrm{H.c.}) +J \sum_j \tilde{\mathbf{S}}_{j}\tilde{\mathbf{S}}_{j+1}\cdot(1-h_j^\dagger h_j)
\end{eqnarray*}
for a single hole in the chain.

The generalization to multiple holes is straightforward. Additionally to considering only spinons on occupied sites, we explicitly map the chargons to the bonds of the squeezed spin chain by introducing new operators $\tilde{h}_{j+1/2}=h_{\tilde{j}}$ where $\tilde{j}=\sum_{i \leq j}(1-h_i^\dagger h_i)$, such that all adjacent chargons are mapped to the same link. For a chain of length $L$ with $N$ particles, the basis states are then of the form
\begin{eqnarray}
    \prod_{j=j_1,...,j_{L-N}} \tilde{h}^\dagger_{j+1/2}|0\rangle \otimes \prod_{j=1}^N \tilde{f}^\dagger_{j,\alpha(j)}\underbrace{|0\rangle \otimes ... \otimes |0\rangle}_{N\text{ times}}. \hspace{3mm}
\end{eqnarray}
Using the same steps as in \req{\ref{eq:kinetic-energy-squeezed}} again results in the same form of the kinetic energy. The spin part has to be modified slightly, since interactions are switched off when there is at least one chargon present on the link. This is achieved by exchanging $(1-h_j^\dagger h_j) \rightarrow (1-\Theta (\tilde{h}_{j+1/2}^\dagger \tilde{h}_{j+1/2}))$ in \req{\ref{eq:spins-squeezed}}. Here, $\Theta(\cdot)$ denotes the step function and we require $\Theta(0)=0$. To treat the density-density interactions in \req{\ref{eq:tJ-model}} we first note that a single hole in the original model switches off the interactions on two adjacent sites. More generally, $k>0$ adjacent holes  switch off $k+1$ bonds. Hence, the number of deactivated density-density interactions are related to the number of chargons on each bond via 
\begin{eqnarray}
    \mathcal{D} = \frac{1}{4} \bigg[ L - \sum_j \kappa(\tilde{h}_{j+1/2}^\dagger \tilde{h}_{j+1/2}) \bigg ]
\end{eqnarray}
where we defined 
\begin{eqnarray}
    \kappa(x) 
    \begin{cases}
    0 \; \; \; \; \; \; \; \; \, x=0 \\
    x+1 \; \; \text{else}
    \end{cases}.
\end{eqnarray}
One then notes that the kinetic energy and density interactions depend only on the chargons. Additionally, the spin term may be split into a Heisenberg term coupling only to the spinons and an interaction term
\begin{eqnarray*}
    \mathcal{S} &= J \sum_j \tilde{\mathbf{S}}_{j}\tilde{\mathbf{S}}_{j+1}-\underbrace{J \sum_j \tilde{\mathbf{S}}_{j}\tilde{\mathbf{S}}_{j+1} \Theta(\tilde{h}_{j+1/2}^\dagger \tilde{h}_{j+1/2})}_{H_{\text{int}}}.
\end{eqnarray*}
In summary, this results in \req{\ref{eq:tJ-model-squeezed-space}} and \req{\ref{eq:tJ-model-description}} of the main text. 

In order to express the staggered potential from \req{\ref{eq:staggered-potential-drive}} in squeezed space, let us first recall that only occupied sites are considered in squeezed space. This implies that the prefactor
\begin{eqnarray}
    (-1)^{j} \xrightarrow{\text{occupied } j} (-1)^{j+\sum_{k < j}\tilde{h}^\dagger_{k+1/2}\tilde{h}_{k+1/2}}
    \label{eq:staggered-prefactor-squeezed}
\end{eqnarray}
aquires a nonlocal form in terms of the chargon operators in squeezed space. Additionally, the number of chargons on a bond $j+1/2$ describes the number of consecutive holes in the chain. For any even number of concecutive holes, the contributions due to the staggered potential cancel, leading to a net contribution of zero. We thus only aquire a contribution of $+1$ when the occupation number on bond $j+1/2$ is odd, which is achieved by a factor $(1-(-1)^{\tilde{h}^\dagger_{j+1/2}\tilde{h}_{j+1/2}})/2$. This leads to the form
\begin{widetext}
\begin{eqnarray}
    \label{eq:staggered-potential-squeezed}
    \hat{V} = \sum_j (-1)^j n_j \rightarrow \frac{1}{2}\sum_{\text{occupied } j} (-1)^{j+\sum_{k < j}\tilde{h}^\dagger_{k+1/2}\tilde{h},_{k+1/2}}\cdot(1-(-1)^{\tilde{h}^\dagger_{j+1/2}\tilde{h}_{j+1/2}}) + \text{const.} \nonumber
\end{eqnarray}
of the staggered potential introduced in \req{\ref{eq:staggered-potential-drive}}.

Using \req{\ref{eq:staggered-prefactor-squeezed}} and previous results in this section, the drive from \req{\ref{eq:staggered-hopping-drive}} induced by driving the Hubbard model with a staggered potential is expressed in squeezed space as follows: The staggered hopping term becomes
\begin{eqnarray}
     gf(\tau)\mathcal{P}_0 \bigg[ \sum_{j,\alpha=\uparrow, \downarrow}(-1)^j c^\dagger_{j,\alpha}c_{j+1, \alpha} + \mathrm{H.c.}) \bigg] \mathcal{P}_0 \rightarrow gf(\tau)\sum_{j} (-1)^{j+\sum_{k < j}(\tilde{h}_{k+1/2}^\dagger \tilde{h}_{k+1/2})}(\tilde{h}^\dagger_{j+1/2} \tilde{h}_{j+3/2} +\mathrm{H.c.}). 
\end{eqnarray}
To treat the spin and density terms, we first note that the contribution proportional to $f(\tau)^2$ merely changes $J \rightarrow J(\tau)=J(1+g^2/t^2f(\tau^2))$. The remaining spin and density interactions transform as
\begin{eqnarray}
    \begin{aligned}    
    \frac{2gJ}{t}f(\tau)\sum_j (-1)^j\bigg(\frac{n_{j}n_{j+1}}{4} \bigg) &\rightarrow \frac{gJ}{4t} f(\tau) \sum_j (-1)^{j+\sum_{k<j}(\tilde{h}_{k+1/2}^\dagger \tilde{h}_{k+1/2})}\cdot ( 1+(-1)^{\tilde{h}_{j+1/2}^\dagger \tilde{h}_{j+1/2})}) \\
     -\frac{2gJ}{t}f(\tau) \sum_j (-1)^j \bigg( \mathbf{S}_{j}\mathbf{S}_{j+1} \bigg) &\rightarrow -\frac{2gJ}{t}f(\tau) \sum_j (-1)^{j+\sum_{k < j}(\tilde{h}_{k+1/2}^\dagger \tilde{h}_{k+1/2})} \cdot \tilde{\mathbf{S}}_{j}\tilde{\mathbf{S}}_{j+1}\cdot(1-\Theta(\tilde{h}_{j+1/2}^\dagger \tilde{h}_{j+1/2})). \hspace{5mm}
     \end{aligned}
\end{eqnarray}
Here, all sums on the right side only run over all occupied sites $j$ and the density term is understood in similar fashion as Eq. \eqref{eq:staggered-potential-squeezed}, with only even numbers of adjacent holes causing a nonzero contribution. 
\end{widetext}

\section{Additional numerical data}
\label{A:additional-data}

In this section, we provide additional numerical data, as well as an extended discussion on introducing randomness into the drive. In particular, we emphasize that this approach can be challenging to control in a general setting.

\subsection{Class (III): Drive with strong spin coupling}

\begin{figure}[h]
    \centering
    \includegraphics{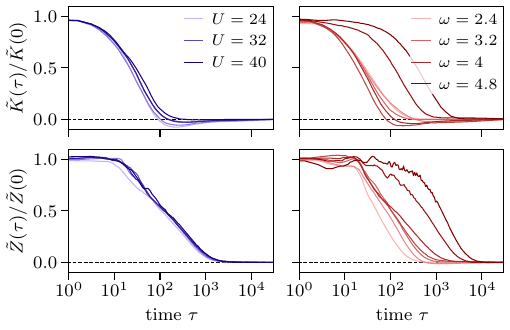}
    \caption{\textbf{Breakdown of fractionalized prethermalization:} A drive that strongly couples to both quasiparticles leads to a breakdown of the fractionalized prethermal plateau. Consequently, heating times of the charge and spin sector agree and any dependence on $U$ (left) as well as frequency $\omega$ (right) is absent up to frequencies $\omega \approx 4.4t$ at which the whole system starts to prethermalize.}
    \label{fig:A6-counter-example}
\end{figure}

In Sec. \ref{sec:fractionalized-prethermalization} of the main text, we have analyzed different drive protocols in which a clear fractionalized prethermal regime is observable. Here, we introduce a drive protocol that couples strongly to the spinons. To this end, we drive the tJ-model with a next-nearest neighbor spin interaction
\begin{eqnarray}
    V = \sum_j S_j^zS_{j+2}^z.
\end{eqnarray}
For this type of drive, a quick heating of spin and charge degrees of freedom is observable; see \rfig{\ref{fig:A6-counter-example}}. Specifically, the strong dependence on onsite interaction $U$ and frequency $\omega$ at intermediate frequencies is not present anymore, demonstrating breakdown of fractionalized prethermalization.

\subsection{Finite size effects in numerical simulations}

\begin{figure}[h]
    \centering
    \includegraphics{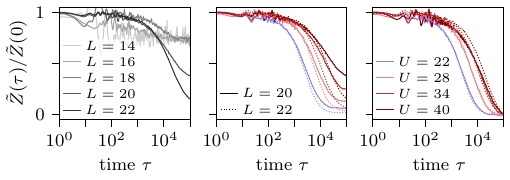}
    \caption{\textbf{Finite size effects:} The correct qualitative behavior of the spin-correlations emerges for system sizes $L\geq20$ (left). While data for $L=20$ and $L=22$ is in good agreement for small onsite interactions $U$, at the largest evaluated values we still encounter considerable finite size effects, manifesting in the correlations not relaxing to their true infinite temperature value (middle). We circumvent this problem by rescaling with respect to the late time steady state value instead of the infinite temperature value (right), which leads to good agreement of the two system sizes over the full range of onsite interactions $10\leq U \leq 40$.}
    \label{fig:A2-finite-size}
\end{figure}

In the main text, we restrict to simulating systems with $L=20$ sites and $L-N=3$ holes for numerical feasibility. Here we give further remarks on the finite size effects that appear in the spin correlations due to the product structure of the Hilbert space; see \rfig{\ref{fig:A2-finite-size}}. First, we note that the correct qualitative behavior emerges only for system sizes $L\geq20$ (with three holes in the system). Additionally, the spin correlations do not relax to their true infinite temperature value but instead start to fluctuate around a small but finite value that decreases with increasing system size. This effect is more prominent for large onsite interactions $U$, hosting stronger spin-charge separation. We circumvent this problem by rescaling the spin correlations with their late time steady-state value $\tilde{Z}(\tau) = Z(\tau) - Z(\tau=10^5t)$ instead of the infinite temperature value $\mathrm{Tr}(Z)/\mathrm{Tr}(\mathcal{I})$. This leads to excellent agreement between data for $L=20$ and $L=22$ over most of the evaluated range of onsite interactions $10\leq U\leq 40$ with some small deviations appearing at larger $U$, see \rfig{\ref{fig:A2-finite-size}}. Subtracting the late-time value is additionally justified by further noting that the late time steady state for the largest system size $L=22$ is very close to an infinite temperature state.

\subsection{Effect of noise on the lifetime}

The driven tJ-model introduced in \req{\ref{eq:tJ-model-with-drive}} is generally expected to thermalize to an infinite temperature state at late times. In exact diagonalization studies, finite size effects can however cause the dynamics to not be fully ergodic, see \rfig{\ref{fig:A2-finite-size}}. The consequence is that observables do not relax to their infinite temperature expectation values. Previous works \cite{fleckenstein2021prethermalization, fleckenstein2021thermalization} have demonstrated that applying a small random perturbation $T\rightarrow T(1+\delta)$ to the period of the drive in each cycle can restore ergodicity by introducing small perturbations to the eigenstates of the Floquet Hamiltonian. In the high frequency limit, this approach only causes multiplicative corrections to the terms in the Magnus expansion and preserves the exponential scaling for sufficiently small $\delta$. Generally, the dynamics of an arbitrary system with statistical periodicity is however governed by an effective Lindblad equation \cite{sieberer2018statistical}. Therefore, even small amounts of randomness $\delta$ induce an additional relaxation channel. Applying Matthiessen's rule to the driven tJ-model investigated throughout this work motivates the Ansatz
\begin{eqnarray}
    \frac{1}{\tau_\mathrm{th}} = \frac{1}{\tau_U}+\frac{1}{\tau_\delta}
    \label{eq:matthiessen}
\end{eqnarray}
for the thermalization time $\tau_\mathrm{th}$ of the spin sector. Here, $\tau_U$ refers to the intrinsic lifetime of the model and $\tau_\delta$ to the relaxation time induced by adding a small amount of randomness to the period of the drive. From the above equation it is immediately apparent that the method is well controlled as long as the relaxation time induced by adding randomness $\tau_\delta \gg \tau_U$ is much larger than the intrisic thermalization time of the system.

\begin{figure}
    \centering
    \includegraphics{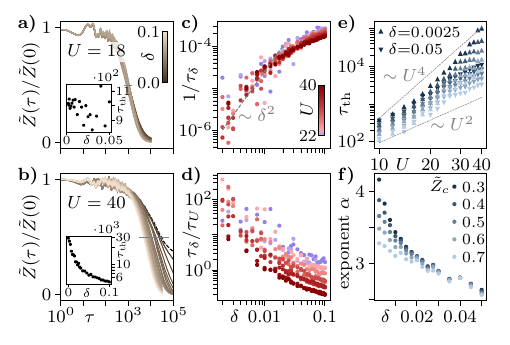}
    \caption{\textbf{Effect of statistical noise on the lifetime:} (a) and (b): Dynamics of the tJ-model driven with a staggered potential for different onsite interaction $U$. The effect of statistical randomness $\delta$ is more prominent for large $U$. The relaxation time $\tau_\delta$ follows a FGR scaling at small $\delta$ as predicted by \cite{sieberer2018statistical} (c). The ratio $\tau_\delta/\tau_U$ of randomness induced lifetime and intrinsic lifetime is already on the order $\mathcal{O}(1)$ for large onsite interactions $U$ and $\delta\sim 0.005$ (d), indicating that the approximation is not well controlled in the present setting. This becomes even more apparent when comparing the resulting lifetimes of the fractional prethermal plateau at different randomness strength (e) and fitting the power law exponent $\tau_\mathrm{th} \sim U^\alpha$ as function of $\delta$ (f) determined at different values $\tilde Z_c$ of $\tilde Z (\tau)/\tilde Z(0)$. It is apparent, that the correct value is approached only in the regime where the lifetime of the fractional prethermal plateau is governed predominantly by randomness. }
    \label{fig:A3-randomness}
\end{figure}

In \rfig{\ref{fig:A3-randomness}}(a) and (b) we show numerical results for the dynamics generated by the tJ-model driven with the staggered potential introduced in \req{\ref{eq:staggered-potential-drive}} and uniform random noise applied in each period. We first note that introducing randomness has minimal effect for small onsite interaction $U$ while decreasing the lifetime up to a factor of five for a randomness level $\delta=0.1$ at large $U=40$, already indicating that the approach is not well controlled in our setting. As before, we define the lifetime $\tau_\mathrm{th}$ as the time where the spin correlations have lost half of their initial value $\tilde{Z}(\tau)/\tilde{Z}(0) = 0.5$. Let us emphasize that contrarily to \rfig{\ref{fig:staggered-pot-drive}}, here we rescale the spin coupling with respect to the infinite temperature value, as the randomness restores ergodicity of the dynamics.

To gain better insight, we use \req{\ref{eq:matthiessen}} to gain an estimate for the lifetime $1/\tau_\delta = 1/\tau_\mathrm{th}(\delta)-1/\tau_\mathrm{th}(\delta=0)$ where the latter should be seen as the intrinsic lifetime of the model $\tau_U$. As shown in \rfig{\ref{fig:A3-randomness}}(c) and (d), the lifetime $\tau_\delta \sim 1/\delta^2$ is consistent with a Fermi-Golden rule scaling predicted by \cite{sieberer2018statistical} for small $\delta$. Additionally, we observe that the ratio $\tau_\delta/\tau_U \sim \mathcal{O}(1)$ is already of order one for large $U$ and randomness $\delta \sim 0.005$, demonstrating that adding even a small amount of randomness overwhelms  the lifetime of the fractional plateau and the resulting lifetime $\tau_\mathrm{th}$ is strongly influenced by $\tau_\delta$. This is further demonstrated in \rfig{\ref{fig:A3-randomness}}(e) and (f) where we approximate the exponent $\alpha$ governing the power law $\tau_\mathrm{th}(\delta) \sim U^{\alpha(\delta)}$. The correct exponent $\alpha=2$ is approached only at large $\delta>0.05$ where the lifetime is however strongly dependent on the introduced randomness. Furthermore, for small $\delta$, exponents are incorrect due to the generated dynamics being close to the unperturbed model, for which the spin energy relaxes to different late time values not accounted for within this analysis.

To summarize, our analysis demonstrates that while adding statistical randomness to a periodically driven system can successfully restore ergodicity, it generally does not provide a well-controlled approximation. The underlying reason is that in a general setting, adding randomness restores ergodicity on relevant timescales only when the relaxation time induced by the randomness $\tau_\delta \sim \tau_U$ is on the order of the intrinsic lifetime of the system. The success e.g. in the prethermal limit however relies on the fact, that adding uniform randomness in each period merely renormalizes the resulting effective Hamiltonian, but still keeps the general structure of the high frequency expansion intact.

\section{Schrieffer-Wolff transformation of the staggered hopping}
\label{A:Schrieffer-Wolff}

In this appendix, we show additional details for the modulated staggered hopping discussed in Sec.$\,$\ref{sec:Floquet-expansion-Hubbard}. We use the same notation as in the main text and denote the driven Hubbard Hamiltonian by
\begin{eqnarray}
    \begin{aligned}
        H_0 &= \frac{U}{2}\sum_j n_j(n_j-1) \\
        t\mathcal{V}(\tau) &= -t\sum_{j,\alpha=\uparrow,\downarrow} A_{j,\alpha} c^\dagger_{i,\alpha}c_{j+1,\alpha} + \mathrm{H.c.}.
    \end{aligned}
\end{eqnarray}
where $A_{j,\alpha} = (1-\frac{g}{t}f(\tau)(-1)^j)$ and $t\mathcal{V}(\tau)$ accounts for the total perturbation to $H_0$ consisting of the static hopping and a the time dependent staggered hopping
\begin{eqnarray}
    gf(\tau)V = g f(\tau)\sum_{j,\alpha=\uparrow, \downarrow}(-1)^j(c^\dagger_{j,\alpha}c_{j+1, \alpha} + \mathrm{H.c.})
\end{eqnarray}
Our goal is to find the low energy effective theory via the time dependent Schrieffer-Wolff rotation introduced in Sec.$\,$\ref{sec:Floquet-expansion-Hubbard}. Hence, we need to determine the components $S_n(\tau)$ of the generator $S(\tau) = \sum_n \epsilon^n S_n(\tau)$ recursively using \req{\ref{eq:time-dependent-SW-general-condition}} by requiring that the offdiagonal terms vanish. As in the main text, we compute the expansion with $t$ as small parameter $\epsilon$ and compute the remaining terms from \req{\ref{eq:Heff-time-dependent}} up to second order in $t$ to obtain the effective Hamiltonian. The consituting equation for $S_1(\tau)$ is given in \req{\ref{eq:time-dependent-SW-1st-order}} and reads
\begin{eqnarray}
    i\partial_\tau S_1(\tau) + \mathcal{V}(\tau) + [S_1(\tau),H_0] = 0.
\end{eqnarray}
We have to eliminate the offdiagonal terms 
\begin{eqnarray}
    i\partial_\tau S_1(\tau)+\mathcal{P}_0 \mathcal{V}(\tau) \mathcal{P}_c+\mathcal{P}_c \mathcal{V}(\tau) \mathcal{P}_0 +[S_1(\tau),H_0] \hspace{3mm}
    \label{eq:app-SW-offdiag-general}
\end{eqnarray}
where $\mathcal{P}_0$ again defines the projector onto the subspace with at most one particle per site and $\mathcal{P}_c$ its complement.  

We first solve \req{\ref{eq:app-SW-offdiag-general}} ignoring the derivative, equivalent to a time independent transformation with modified interaction strength $t \rightarrow tA_{j,\alpha}$. The offdiagonal terms are eliminated by choosing
\begin{eqnarray}
    S_1(\tau) = \frac{1}{U}[\mathcal{P}_c\mathcal{V}(\tau)\mathcal{P}_0-\mathcal{P}_0\mathcal{V}(\tau)\mathcal{P}_c].
    \label{eq:app-SW-S1-ansatz-base}
\end{eqnarray}

And the second order contribution to the effective Hamiltonian is determined from the diagonal contributions arising from the Baker-Campbell-Hausdorff expansion  
\begin{eqnarray}
    [S_1,\mathcal{P}_0 \mathcal{V} \mathcal{P}_c+\mathcal{P}_c \mathcal{V} \mathcal{P}_0]+\frac{1}{2}[S_1,[S_1,H_0]],
    \label{eq:app-SW-static-second-order-base}
\end{eqnarray}
which result in a total contribution
\begin{eqnarray}
    H_\text{eff}^{(2)} = - \frac{t^2}{U} \mathcal{P}_0 V \mathcal{P}_c V \mathcal{P}_0
    \label{eq:second-order-contribution}
\end{eqnarray}
to the lowest subspace. They account for second order exchange processes of the form 
\begin{eqnarray}
    |\bar{\alpha},\alpha\rangle \rightarrow |2,0\rangle \rightarrow 
    \begin{cases}
        |\bar{\alpha}, \alpha \rangle \\
        |\alpha, \bar{\alpha} \rangle
    \end{cases}
\end{eqnarray}
similar to the Hubbard model but with modified strength $A_{j,\alpha}^2$. This results in the same form of an effective spin-spin interaction but with strength $J' = J\cdot A_{j,\alpha}^2$. Here, we again neglect processes of the form $|0,\bar{\alpha},\alpha\rangle \rightarrow |0,2,0\rangle \rightarrow |\alpha,\bar{\alpha},0\rangle$, where the particles hop over, suppressed by a factor $1/8$.

Next, we need to determine the influence of the derivative on this result. We make the same ansatz for $S_1(\tau)$ as in \req{\ref{eq:app-SW-S1-ansatz-base}} but we do not fix the coefficients to be the same as in $\mathcal{V}$ and explicitly write it with hermitian conjugate. That is, 
\begin{eqnarray}
    \tilde{S}_1(\tau) &=& \frac{1}{U} \mathcal{P}_c \Big\{ -\sum_{j,\alpha} B_{j,\alpha}(\tau) c^\dagger_{j,\alpha}c_{j+1, \alpha}+\mathrm{H.c.} \Big\} \mathcal{P}_0 -\mathrm{H.c.} \nonumber\\&\equiv& \frac{1}{U}\mathcal{P}_c \mathcal{V}_s \mathcal{P}_0 - \mathrm{H.c.}
    \label{eq:app-SW-S1-tilde-base}
\end{eqnarray}
where $B_{j,\alpha}(\tau)$ is not necessarily real. The commutator $[\tilde{S}_1(\tau), H_0]$ is the same as in the case before, and the resulting equation becomes
\begin{eqnarray}
    \frac{i}{U}\partial_\tau \mathcal{P}_c \mathcal{V}_s \mathcal{P}_0 + \mathcal{P}_c \mathcal{V}(\tau) \mathcal{P}_0 - \mathcal{P}_c \mathcal{V}_s \mathcal{P}_0 = 0
\end{eqnarray}
which leads to the following differential equations for the coefficients:
\begin{eqnarray}
    -\frac{i}{U}\partial_\tau B_{j,\alpha}(\tau) - 1 + (-1)^jg f(\tau) + B_{j,\alpha}(\tau) = 0 \hspace{6mm}
\end{eqnarray}
The general solution of this equation for arbitrary $f(\tau)$ cannot be determined analytically; therefore, we consider $f(\tau)=\sin(\omega\tau)$ as an example. In this case, a  particular solution is given by
\begin{eqnarray}
    B_{j,\alpha}(\tau )&=&1 - (-1)^j \Big( \frac{g U^2\sin(\omega\tau)}{U^2-\omega^2}+\frac{i g U\omega\cos(\omega\tau)}{U^2-\omega^2} \Big) \hspace{9mm}
\end{eqnarray}
and we may set the coefficient of the general solution to zero since we are only interested in a particular solution.

Considering the derivative, the second order effective Hamiltonian consists of two contributions: (1) A term
\begin{eqnarray}
    H_\partial^{(2)} = \frac{1}{2} \big[S_1(\tau)\partial_\tau S_1(\tau)-(\partial_\tau S_1(\tau))S_1(\tau) \big ]
     \label{eq:app-SW-dynamic-second-order-base}
\end{eqnarray}
arising from the derivative, which results in subleading contributions of order $\mathcal{O}(\omega t^2/U^2)$ to the effective spin interaction. (2) The static part in \req{\ref{eq:app-SW-static-second-order-base}} with modified coefficients: Instead of \req{\ref{eq:second-order-contribution}}, the resulting contribution in the subspace $\mathcal{P}_0$ is
    \begin{eqnarray}
        -\frac{1}{U} ( \mathcal{P}_0 \mathcal{V} \mathcal{P}_c \mathcal{V}_s \mathcal{P}_0 + \mathcal{P}_0 \mathcal{V}_s^\dagger \mathcal{P}_c \mathcal{V} \mathcal{P}_0 ) + \frac{1}{U} (\mathcal{P}_0 \mathcal{V}_s^\dagger\mathcal{P}_c \mathcal{V}_s\mathcal{P}_0). \hspace{9mm}
    \end{eqnarray}
Following similar steps as above, we obtain for the prefactor $J'$
    \begin{eqnarray}
        J'/J &=& 2\text{Re}(B_{j,\alpha})A_{j,\alpha}-|B_{j,\alpha}|^2 = \\ \nonumber
&&2\bigg( 1-(-1)^j\frac{gU^2\sin(\omega\tau)}{U^2-\omega^2}\bigg)(1-(-1)^jg\sin(\omega\tau))-\\ \nonumber
        &&\bigg( 1-(-1)^j\frac{gU^2\sin(\omega\tau)}{U^2-\omega^2}\bigg)^2- \frac{g^2U^2\omega^2\cos(\omega\tau)^2}{(U^2-\omega^2)^2}, 
    \end{eqnarray}
where the last part is the contribution due to additional terms coming from the derivative.

\section{Optical lattice realization of the driven Hubbard model}
\label{A:single particle SE}

\begin{figure}[h]
    \centering
    \includegraphics{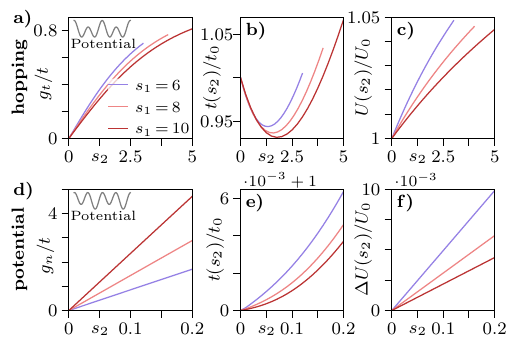}
    \caption{\textbf{Lattice Hamiltonian parameters of the optical lattice setup:} (a-c) Staggered hopping amplitude, average hopping constant and onsite interaction strength as function of the superlattice strength $s_2$ for \textbf{Case I}. (d-f) Staggered onsite potential, hopping constant and staggering of the onsite interaction strength for \textbf{Case II}.}
    \label{fig:A5-lattice-parameters}
\end{figure}

Here, we numerically solve the band structures with superlattices that can be used to modulate the hopping amplitudes and the staggered potential. Specifically, we consider the two optical potentials introduced in \req{\ref{eq:lattice-potential}}. Both potentials have a two site unit cell $a=\frac{2\pi}{k_L}$ and the relevant physics in the limit of deep lattices are captured by the two lowest bands. First, we numerically solve the singe-particle Schrödinger equation using standard techniques (see e.g. \cite{bissbort2013dynamical}, Chapter 2) to obtain the single particle band structure, and fit it with the respective tight binding model to obtain the parameters of the quadratic Hamiltonian. We then construct the maximally localized Wannier functions (MLWF) $w_{j}(x) = \langle x|j\rangle$ as eigenstates of the position operator projected onto the two lowest bands, allowing contributions from both bands, see \cite{bissbort2013dynamical}. Using the MLWF we compute the onsite interaction strength as
\begin{eqnarray}
    U_l^\alpha = \frac{4\pi\hbar^2a_s}{2m} \int dx |w_{l,\alpha}(x)|^4
\end{eqnarray}
with $a_s$ the s-wave scattering length and $m$ the mass of the atoms. We consider the following cases:

\noindent \textbf{Case I:} 
\begin{eqnarray}
     V(x) &= E_r \bigl ( s_1\sin^2(k_Lx)+s_2\sin^2(\frac{k_L}{2}x+\frac{\pi}{4}) \bigr )
\end{eqnarray}
with lattice Hamiltonian

\begin{eqnarray}
    \begin{aligned}
    H = &\sum_{j,\alpha}(-t+(-1)^jg_t)(c^\dagger_{j,\alpha}c_{j+1, \alpha} + \mathrm{H.c.}) \\
    &+ U\sum_j n_{j,\uparrow}n_{j,\downarrow}
    \end{aligned}
\end{eqnarray}
and tight binding dispersion
\begin{eqnarray}
    \begin{aligned}
    \epsilon_{\text{hop}}(k) &= \pm \sqrt{2t^2+2g^2+2(t^2-g^2)\cos(2k)}.
    \end{aligned}
\end{eqnarray}
This potential implements the driven Hubbard model in \req{\ref{eq:staggered-hopping-drive}} upon modulating $s_2$. In order to realize a strong drive $g$, the strength $s_2$ is considerable. This renormalizes the hopping $t$ and onsite interaction strength $U$. The effect can however be counteracted by dynamically adjusting $s_1(\tau) = s_1 + s_2(\tau)/2.7$, with a numerically determined factor $2.7$ that keeps $t, U$ constant within approximately 5\% relative deviation. We show results in \rfig{\ref{fig:A5-lattice-parameters}} (a-c).
\noindent \textbf{Case II:}
\begin{eqnarray}
     V(x) &= E_r \bigl ( s_1\sin^2(k_Lx)+s_2\sin^2(\frac{k_L}{2}x) \bigr )
\end{eqnarray}
with lattice Hamiltonian
\begin{eqnarray}
    \begin{aligned}
    H = &-t\sum_{j,\alpha} (c^\dagger_{j,\alpha}c_{j+1, \alpha} + \mathrm{H.c.}) + U\sum_j n_{j,\uparrow}n_{j,\downarrow} \\
    &+g_n\sum_j (-1)^j(n_{j,\uparrow}+n_{j,\downarrow})
    \end{aligned}
\end{eqnarray}
and tight binding dispersion
\begin{eqnarray}
    \begin{aligned}
    \epsilon_{\text{pot}}(k) &= \pm \sqrt{g^2+4t^2\cos^2(k)}.
    \end{aligned}
\end{eqnarray}
This potential implements the driven Hubbard model in \req{\ref{eq:staggered-potential-drive}} upon modulating $s_2$. Note that the potential introduces a slight staggering $\Delta U = U_j-U_{j+1}$. Numerical results are shown in \rfig{\ref{fig:A5-lattice-parameters}} (d-f).

\bibliography{biblio}

\begin{thebibliography}{68}%
\makeatletter
\providecommand \@ifxundefined [1]{%
 \@ifx{#1\undefined}
}%
\providecommand \@ifnum [1]{%
 \ifnum #1\expandafter \@firstoftwo
 \else \expandafter \@secondoftwo
 \fi
}%
\providecommand \@ifx [1]{%
 \ifx #1\expandafter \@firstoftwo
 \else \expandafter \@secondoftwo
 \fi
}%
\providecommand \natexlab [1]{#1}%
\providecommand \enquote  [1]{``#1''}%
\providecommand \bibnamefont  [1]{#1}%
\providecommand \bibfnamefont [1]{#1}%
\providecommand \citenamefont [1]{#1}%
\providecommand \href@noop [0]{\@secondoftwo}%
\providecommand \href [0]{\begingroup \@sanitize@url \@href}%
\providecommand \@href[1]{\@@startlink{#1}\@@href}%
\providecommand \@@href[1]{\endgroup#1\@@endlink}%
\providecommand \@sanitize@url [0]{\catcode `\\12\catcode `\$12\catcode
  `\&12\catcode `\#12\catcode `\^12\catcode `\_12\catcode `\%12\relax}%
\providecommand \@@startlink[1]{}%
\providecommand \@@endlink[0]{}%
\providecommand \url  [0]{\begingroup\@sanitize@url \@url }%
\providecommand \@url [1]{\endgroup\@href {#1}{\urlprefix }}%
\providecommand \urlprefix  [0]{URL }%
\providecommand \Eprint [0]{\href }%
\providecommand \doibase [0]{https://doi.org/}%
\providecommand \selectlanguage [0]{\@gobble}%
\providecommand \bibinfo  [0]{\@secondoftwo}%
\providecommand \bibfield  [0]{\@secondoftwo}%
\providecommand \translation [1]{[#1]}%
\providecommand \BibitemOpen [0]{}%
\providecommand \bibitemStop [0]{}%
\providecommand \bibitemNoStop [0]{.\EOS\space}%
\providecommand \EOS [0]{\spacefactor3000\relax}%
\providecommand \BibitemShut  [1]{\csname bibitem#1\endcsname}%
\let\auto@bib@innerbib\@empty
\bibitem [{\citenamefont {Oka}\ and\ \citenamefont
  {Aoki}(2009)}]{oka2009photovoltaic}%
  \BibitemOpen
  \bibfield  {author} {\bibinfo {author} {\bibfnamefont {T.}~\bibnamefont
  {Oka}}\ and\ \bibinfo {author} {\bibfnamefont {H.}~\bibnamefont {Aoki}},\
  }\bibfield  {title} {\bibinfo {title} {Photovoltaic hall effect in
  graphene},\ }\href {https://doi.org/10.1103/PhysRevB.79.081406} {\bibfield
  {journal} {\bibinfo  {journal} {Phys. Rev. B}\ }\textbf {\bibinfo {volume}
  {79}},\ \bibinfo {pages} {081406} (\bibinfo {year} {2009})}\BibitemShut
  {NoStop}%
\bibitem [{\citenamefont {Lindner}\ \emph {et~al.}(2011)\citenamefont
  {Lindner}, \citenamefont {Refael},\ and\ \citenamefont
  {Galitski}}]{lindner2011floquet}%
  \BibitemOpen
  \bibfield  {author} {\bibinfo {author} {\bibfnamefont {N.~H.}\ \bibnamefont
  {Lindner}}, \bibinfo {author} {\bibfnamefont {G.}~\bibnamefont {Refael}},\
  and\ \bibinfo {author} {\bibfnamefont {V.}~\bibnamefont {Galitski}},\
  }\bibfield  {title} {\bibinfo {title} {Floquet topological insulator in
  semiconductor quantum wells},\ }\href
  {https://doi.org/https://doi.org/10.1038/nphys1926} {\bibfield  {journal}
  {\bibinfo  {journal} {Nature Physics}\ }\textbf {\bibinfo {volume} {7}},\
  \bibinfo {pages} {490} (\bibinfo {year} {2011})}\BibitemShut {NoStop}%
\bibitem [{\citenamefont {Goldman}\ and\ \citenamefont
  {Dalibard}(2014)}]{goldman2014periodically}%
  \BibitemOpen
  \bibfield  {author} {\bibinfo {author} {\bibfnamefont {N.}~\bibnamefont
  {Goldman}}\ and\ \bibinfo {author} {\bibfnamefont {J.}~\bibnamefont
  {Dalibard}},\ }\bibfield  {title} {\bibinfo {title} {Periodically driven
  quantum systems: Effective hamiltonians and engineered gauge fields},\ }\href
  {https://doi.org/10.1103/PhysRevX.4.031027} {\bibfield  {journal} {\bibinfo
  {journal} {Phys. Rev. X}\ }\textbf {\bibinfo {volume} {4}},\ \bibinfo {pages}
  {031027} (\bibinfo {year} {2014})}\BibitemShut {NoStop}%
\bibitem [{\citenamefont {Bukov}\ \emph
  {et~al.}(2015{\natexlab{a}})\citenamefont {Bukov}, \citenamefont
  {Gopalakrishnan}, \citenamefont {Knap},\ and\ \citenamefont
  {Demler}}]{bukov2015prethermal}%
  \BibitemOpen
  \bibfield  {author} {\bibinfo {author} {\bibfnamefont {M.}~\bibnamefont
  {Bukov}}, \bibinfo {author} {\bibfnamefont {S.}~\bibnamefont
  {Gopalakrishnan}}, \bibinfo {author} {\bibfnamefont {M.}~\bibnamefont
  {Knap}},\ and\ \bibinfo {author} {\bibfnamefont {E.}~\bibnamefont {Demler}},\
  }\bibfield  {title} {\bibinfo {title} {Prethermal floquet steady states and
  instabilities in the periodically driven, weakly interacting bose-hubbard
  model},\ }\bibfield  {journal} {\bibinfo  {journal} {Physical Review
  Letters}\ }\textbf {\bibinfo {volume} {115}},\ \href
  {https://doi.org/10.1103/physrevlett.115.205301}
  {10.1103/physrevlett.115.205301} (\bibinfo {year}
  {2015}{\natexlab{a}})\BibitemShut {NoStop}%
\bibitem [{\citenamefont {Eckardt}(2017)}]{eckardt2017colloquium}%
  \BibitemOpen
  \bibfield  {author} {\bibinfo {author} {\bibfnamefont {A.}~\bibnamefont
  {Eckardt}},\ }\bibfield  {title} {\bibinfo {title} {Colloquium: Atomic
  quantum gases in periodically driven optical lattices},\ }\bibfield
  {journal} {\bibinfo  {journal} {Reviews of Modern Physics}\ }\textbf
  {\bibinfo {volume} {89}},\ \href
  {https://doi.org/10.1103/revmodphys.89.011004} {10.1103/revmodphys.89.011004}
  (\bibinfo {year} {2017})\BibitemShut {NoStop}%
\bibitem [{\citenamefont {Cooper}\ \emph {et~al.}(2019)\citenamefont {Cooper},
  \citenamefont {Dalibard},\ and\ \citenamefont
  {Spielman}}]{cooper2019topological}%
  \BibitemOpen
  \bibfield  {author} {\bibinfo {author} {\bibfnamefont {N.~R.}\ \bibnamefont
  {Cooper}}, \bibinfo {author} {\bibfnamefont {J.}~\bibnamefont {Dalibard}},\
  and\ \bibinfo {author} {\bibfnamefont {I.~B.}\ \bibnamefont {Spielman}},\
  }\bibfield  {title} {\bibinfo {title} {Topological bands for ultracold
  atoms},\ }\href {https://doi.org/10.1103/RevModPhys.91.015005} {\bibfield
  {journal} {\bibinfo  {journal} {Rev. Mod. Phys.}\ }\textbf {\bibinfo {volume}
  {91}},\ \bibinfo {pages} {015005} (\bibinfo {year} {2019})}\BibitemShut
  {NoStop}%
\bibitem [{\citenamefont {Oka}\ and\ \citenamefont
  {Kitamura}(2019)}]{oka2019floquet}%
  \BibitemOpen
  \bibfield  {author} {\bibinfo {author} {\bibfnamefont {T.}~\bibnamefont
  {Oka}}\ and\ \bibinfo {author} {\bibfnamefont {S.}~\bibnamefont {Kitamura}},\
  }\bibfield  {title} {\bibinfo {title} {Floquet engineering of quantum
  materials},\ }\href
  {https://doi.org/10.1146/annurev-conmatphys-031218-013423} {\bibfield
  {journal} {\bibinfo  {journal} {Annual Review of Condensed Matter Physics}\
  }\textbf {\bibinfo {volume} {10}},\ \bibinfo {pages} {387–408} (\bibinfo
  {year} {2019})}\BibitemShut {NoStop}%
\bibitem [{\citenamefont {Rudner}\ and\ \citenamefont
  {Lindner}(2020)}]{rudner2020band}%
  \BibitemOpen
  \bibfield  {author} {\bibinfo {author} {\bibfnamefont {M.~S.}\ \bibnamefont
  {Rudner}}\ and\ \bibinfo {author} {\bibfnamefont {N.~H.}\ \bibnamefont
  {Lindner}},\ }\bibfield  {title} {\bibinfo {title} {Band structure
  engineering and non-equilibrium dynamics in floquet topological insulators},\
  }\href {https://doi.org/10.1038/s42254-020-0170-z} {\bibfield  {journal}
  {\bibinfo  {journal} {Nature Reviews Physics}\ }\textbf {\bibinfo {volume}
  {2}},\ \bibinfo {pages} {229–244} (\bibinfo {year} {2020})}\BibitemShut
  {NoStop}%
\bibitem [{\citenamefont {Kitagawa}\ \emph {et~al.}(2010)\citenamefont
  {Kitagawa}, \citenamefont {Berg}, \citenamefont {Rudner},\ and\ \citenamefont
  {Demler}}]{kitagawa2010topological}%
  \BibitemOpen
  \bibfield  {author} {\bibinfo {author} {\bibfnamefont {T.}~\bibnamefont
  {Kitagawa}}, \bibinfo {author} {\bibfnamefont {E.}~\bibnamefont {Berg}},
  \bibinfo {author} {\bibfnamefont {M.}~\bibnamefont {Rudner}},\ and\ \bibinfo
  {author} {\bibfnamefont {E.}~\bibnamefont {Demler}},\ }\bibfield  {title}
  {\bibinfo {title} {Topological characterization of periodically driven
  quantum systems},\ }\href {https://doi.org/10.1103/PhysRevB.82.235114}
  {\bibfield  {journal} {\bibinfo  {journal} {Phys. Rev. B}\ }\textbf {\bibinfo
  {volume} {82}},\ \bibinfo {pages} {235114} (\bibinfo {year}
  {2010})}\BibitemShut {NoStop}%
\bibitem [{\citenamefont {Rudner}\ \emph {et~al.}(2013)\citenamefont {Rudner},
  \citenamefont {Lindner}, \citenamefont {Berg},\ and\ \citenamefont
  {Levin}}]{rudner2013anomalous}%
  \BibitemOpen
  \bibfield  {author} {\bibinfo {author} {\bibfnamefont {M.~S.}\ \bibnamefont
  {Rudner}}, \bibinfo {author} {\bibfnamefont {N.~H.}\ \bibnamefont {Lindner}},
  \bibinfo {author} {\bibfnamefont {E.}~\bibnamefont {Berg}},\ and\ \bibinfo
  {author} {\bibfnamefont {M.}~\bibnamefont {Levin}},\ }\bibfield  {title}
  {\bibinfo {title} {Anomalous edge states and the bulk-edge correspondence for
  periodically driven two-dimensional systems},\ }\href
  {https://doi.org/10.1103/PhysRevX.3.031005} {\bibfield  {journal} {\bibinfo
  {journal} {Phys. Rev. X}\ }\textbf {\bibinfo {volume} {3}},\ \bibinfo {pages}
  {031005} (\bibinfo {year} {2013})}\BibitemShut {NoStop}%
\bibitem [{\citenamefont {Roy}\ and\ \citenamefont
  {Harper}(2017)}]{roy2017periodic}%
  \BibitemOpen
  \bibfield  {author} {\bibinfo {author} {\bibfnamefont {R.}~\bibnamefont
  {Roy}}\ and\ \bibinfo {author} {\bibfnamefont {F.}~\bibnamefont {Harper}},\
  }\bibfield  {title} {\bibinfo {title} {Periodic table for floquet topological
  insulators},\ }\href {https://doi.org/10.1103/PhysRevB.96.155118} {\bibfield
  {journal} {\bibinfo  {journal} {Phys. Rev. B}\ }\textbf {\bibinfo {volume}
  {96}},\ \bibinfo {pages} {155118} (\bibinfo {year} {2017})}\BibitemShut
  {NoStop}%
\bibitem [{\citenamefont {Wintersperger}\ \emph {et~al.}(2020)\citenamefont
  {Wintersperger}, \citenamefont {Braun}, \citenamefont {\"{U}nal},
  \citenamefont {Eckardt}, \citenamefont {Liberto}, \citenamefont {Goldman},
  \citenamefont {Bloch},\ and\ \citenamefont
  {Aidelsburger}}]{wintersperger2020realization}%
  \BibitemOpen
  \bibfield  {author} {\bibinfo {author} {\bibfnamefont {K.}~\bibnamefont
  {Wintersperger}}, \bibinfo {author} {\bibfnamefont {C.}~\bibnamefont
  {Braun}}, \bibinfo {author} {\bibfnamefont {F.~N.}\ \bibnamefont {\"{U}nal}},
  \bibinfo {author} {\bibfnamefont {A.}~\bibnamefont {Eckardt}}, \bibinfo
  {author} {\bibfnamefont {M.~D.}\ \bibnamefont {Liberto}}, \bibinfo {author}
  {\bibfnamefont {N.}~\bibnamefont {Goldman}}, \bibinfo {author} {\bibfnamefont
  {I.}~\bibnamefont {Bloch}},\ and\ \bibinfo {author} {\bibfnamefont
  {M.}~\bibnamefont {Aidelsburger}},\ }\bibfield  {title} {\bibinfo {title}
  {Realization of an anomalous floquet topological system with ultracold
  atoms},\ }\href {https://doi.org/10.1038/s41567-020-0949-y} {\bibfield
  {journal} {\bibinfo  {journal} {Nature Physics}\ }\textbf {\bibinfo {volume}
  {16}},\ \bibinfo {pages} {1058–1063} (\bibinfo {year} {2020})}\BibitemShut
  {NoStop}%
\bibitem [{\citenamefont {Braun}\ \emph {et~al.}(2024)\citenamefont {Braun},
  \citenamefont {Saint-Jalm}, \citenamefont {Hesse}, \citenamefont {Arceri},
  \citenamefont {Bloch},\ and\ \citenamefont {Aidelsburger}}]{braun2024real}%
  \BibitemOpen
  \bibfield  {author} {\bibinfo {author} {\bibfnamefont {C.}~\bibnamefont
  {Braun}}, \bibinfo {author} {\bibfnamefont {R.}~\bibnamefont {Saint-Jalm}},
  \bibinfo {author} {\bibfnamefont {A.}~\bibnamefont {Hesse}}, \bibinfo
  {author} {\bibfnamefont {J.}~\bibnamefont {Arceri}}, \bibinfo {author}
  {\bibfnamefont {I.}~\bibnamefont {Bloch}},\ and\ \bibinfo {author}
  {\bibfnamefont {M.}~\bibnamefont {Aidelsburger}},\ }\bibfield  {title}
  {\bibinfo {title} {Real-space detection and manipulation of topological edge
  modes with ultracold atoms},\ }\href
  {https://doi.org/10.1038/s41567-024-02506-z} {\bibfield  {journal} {\bibinfo
  {journal} {Nature Physics}\ }\textbf {\bibinfo {volume} {20}},\ \bibinfo
  {pages} {1306} (\bibinfo {year} {2024})}\BibitemShut {NoStop}%
\bibitem [{\citenamefont {Po}\ \emph {et~al.}(2017)\citenamefont {Po},
  \citenamefont {Fidkowski}, \citenamefont {Vishwanath},\ and\ \citenamefont
  {Potter}}]{po2017radical}%
  \BibitemOpen
  \bibfield  {author} {\bibinfo {author} {\bibfnamefont {H.~C.}\ \bibnamefont
  {Po}}, \bibinfo {author} {\bibfnamefont {L.}~\bibnamefont {Fidkowski}},
  \bibinfo {author} {\bibfnamefont {A.}~\bibnamefont {Vishwanath}},\ and\
  \bibinfo {author} {\bibfnamefont {A.~C.}\ \bibnamefont {Potter}},\ }\bibfield
   {title} {\bibinfo {title} {Radical chiral floquet phases in a periodically
  driven kitaev model and beyond},\ }\href
  {https://doi.org/10.1103/PhysRevB.96.245116} {\bibfield  {journal} {\bibinfo
  {journal} {Phys. Rev. B}\ }\textbf {\bibinfo {volume} {96}},\ \bibinfo
  {pages} {245116} (\bibinfo {year} {2017})}\BibitemShut {NoStop}%
\bibitem [{\citenamefont {Fulga}\ \emph {et~al.}(2019)\citenamefont {Fulga},
  \citenamefont {Maksymenko}, \citenamefont {Rieder}, \citenamefont {Lindner},\
  and\ \citenamefont {Berg}}]{fulga2019topology}%
  \BibitemOpen
  \bibfield  {author} {\bibinfo {author} {\bibfnamefont {I.~C.}\ \bibnamefont
  {Fulga}}, \bibinfo {author} {\bibfnamefont {M.}~\bibnamefont {Maksymenko}},
  \bibinfo {author} {\bibfnamefont {M.~T.}\ \bibnamefont {Rieder}}, \bibinfo
  {author} {\bibfnamefont {N.~H.}\ \bibnamefont {Lindner}},\ and\ \bibinfo
  {author} {\bibfnamefont {E.}~\bibnamefont {Berg}},\ }\bibfield  {title}
  {\bibinfo {title} {Topology and localization of a periodically driven kitaev
  model},\ }\href {https://doi.org/10.1103/PhysRevB.99.235408} {\bibfield
  {journal} {\bibinfo  {journal} {Phys. Rev. B}\ }\textbf {\bibinfo {volume}
  {99}},\ \bibinfo {pages} {235408} (\bibinfo {year} {2019})}\BibitemShut
  {NoStop}%
\bibitem [{\citenamefont {Will}\ \emph {et~al.}(2025)\citenamefont {Will},
  \citenamefont {Cochran}, \citenamefont {Rosenberg}, \citenamefont {Jobst},
  \citenamefont {Eassa}, \citenamefont {Roushan}, \citenamefont {Knap},
  \citenamefont {Gammon-Smith},\ and\ \citenamefont
  {Pollmann}}]{will2025probing}%
  \BibitemOpen
  \bibfield  {author} {\bibinfo {author} {\bibfnamefont {M.}~\bibnamefont
  {Will}}, \bibinfo {author} {\bibfnamefont {T.~A.}\ \bibnamefont {Cochran}},
  \bibinfo {author} {\bibfnamefont {E.}~\bibnamefont {Rosenberg}}, \bibinfo
  {author} {\bibfnamefont {B.}~\bibnamefont {Jobst}}, \bibinfo {author}
  {\bibfnamefont {N.~M.}\ \bibnamefont {Eassa}}, \bibinfo {author}
  {\bibfnamefont {P.}~\bibnamefont {Roushan}}, \bibinfo {author} {\bibfnamefont
  {M.}~\bibnamefont {Knap}}, \bibinfo {author} {\bibfnamefont {A.}~\bibnamefont
  {Gammon-Smith}},\ and\ \bibinfo {author} {\bibfnamefont {F.}~\bibnamefont
  {Pollmann}},\ }\href {https://arxiv.org/abs/2501.18461} {\bibinfo {title}
  {Probing non-equilibrium topological order on a quantum processor}} (\bibinfo
  {year} {2025}),\ \Eprint {https://arxiv.org/abs/2501.18461} {arXiv:2501.18461
  [quant-ph]} \BibitemShut {NoStop}%
\bibitem [{\citenamefont {Bukov}\ \emph
  {et~al.}(2015{\natexlab{b}})\citenamefont {Bukov}, \citenamefont
  {D’Alessio},\ and\ \citenamefont {Polkovnikov}}]{bukov2015universal}%
  \BibitemOpen
  \bibfield  {author} {\bibinfo {author} {\bibfnamefont {M.}~\bibnamefont
  {Bukov}}, \bibinfo {author} {\bibfnamefont {L.}~\bibnamefont {D’Alessio}},\
  and\ \bibinfo {author} {\bibfnamefont {A.}~\bibnamefont {Polkovnikov}},\
  }\bibfield  {title} {\bibinfo {title} {Universal high-frequency behavior of
  periodically driven systems: from dynamical stabilization to floquet
  engineering},\ }\href {https://doi.org/10.1080/00018732.2015.1055918}
  {\bibfield  {journal} {\bibinfo  {journal} {Advances in Physics}\ }\textbf
  {\bibinfo {volume} {64}},\ \bibinfo {pages} {139–226} (\bibinfo {year}
  {2015}{\natexlab{b}})}\BibitemShut {NoStop}%
\bibitem [{\citenamefont {Kuwahara}\ \emph {et~al.}(2016)\citenamefont
  {Kuwahara}, \citenamefont {Mori},\ and\ \citenamefont
  {Saito}}]{kuwahara2016floquet}%
  \BibitemOpen
  \bibfield  {author} {\bibinfo {author} {\bibfnamefont {T.}~\bibnamefont
  {Kuwahara}}, \bibinfo {author} {\bibfnamefont {T.}~\bibnamefont {Mori}},\
  and\ \bibinfo {author} {\bibfnamefont {K.}~\bibnamefont {Saito}},\ }\bibfield
   {title} {\bibinfo {title} {Floquet–magnus theory and generic transient
  dynamics in periodically driven many-body quantum systems},\ }\href
  {https://doi.org/10.1016/j.aop.2016.01.012} {\bibfield  {journal} {\bibinfo
  {journal} {Annals of Physics}\ }\textbf {\bibinfo {volume} {367}},\ \bibinfo
  {pages} {96–124} (\bibinfo {year} {2016})}\BibitemShut {NoStop}%
\bibitem [{\citenamefont {Abanin}\ \emph {et~al.}(2015)\citenamefont {Abanin},
  \citenamefont {De~Roeck},\ and\ \citenamefont
  {Huveneers}}]{abanin2015exponentially}%
  \BibitemOpen
  \bibfield  {author} {\bibinfo {author} {\bibfnamefont {D.~A.}\ \bibnamefont
  {Abanin}}, \bibinfo {author} {\bibfnamefont {W.}~\bibnamefont {De~Roeck}},\
  and\ \bibinfo {author} {\bibfnamefont {F.}~\bibnamefont {Huveneers}},\
  }\bibfield  {title} {\bibinfo {title} {Exponentially slow heating in
  periodically driven many-body systems},\ }\bibfield  {journal} {\bibinfo
  {journal} {Physical Review Letters}\ }\textbf {\bibinfo {volume} {115}},\
  \href {https://doi.org/10.1103/physrevlett.115.256803}
  {10.1103/physrevlett.115.256803} (\bibinfo {year} {2015})\BibitemShut
  {NoStop}%
\bibitem [{\citenamefont {Canovi}\ \emph {et~al.}(2016)\citenamefont {Canovi},
  \citenamefont {Kollar},\ and\ \citenamefont
  {Eckstein}}]{canovi2016stroboscopic}%
  \BibitemOpen
  \bibfield  {author} {\bibinfo {author} {\bibfnamefont {E.}~\bibnamefont
  {Canovi}}, \bibinfo {author} {\bibfnamefont {M.}~\bibnamefont {Kollar}},\
  and\ \bibinfo {author} {\bibfnamefont {M.}~\bibnamefont {Eckstein}},\
  }\bibfield  {title} {\bibinfo {title} {Stroboscopic prethermalization in
  weakly interacting periodically driven systems},\ }\bibfield  {journal}
  {\bibinfo  {journal} {Physical Review E}\ }\textbf {\bibinfo {volume} {93}},\
  \href {https://doi.org/10.1103/physreve.93.012130}
  {10.1103/physreve.93.012130} (\bibinfo {year} {2016})\BibitemShut {NoStop}%
\bibitem [{\citenamefont {Mori}\ \emph {et~al.}(2016)\citenamefont {Mori},
  \citenamefont {Kuwahara},\ and\ \citenamefont {Saito}}]{mori2016rigorous}%
  \BibitemOpen
  \bibfield  {author} {\bibinfo {author} {\bibfnamefont {T.}~\bibnamefont
  {Mori}}, \bibinfo {author} {\bibfnamefont {T.}~\bibnamefont {Kuwahara}},\
  and\ \bibinfo {author} {\bibfnamefont {K.}~\bibnamefont {Saito}},\ }\bibfield
   {title} {\bibinfo {title} {Rigorous bound on energy absorption and generic
  relaxation in periodically driven quantum systems},\ }\href
  {https://doi.org/10.1103/PhysRevLett.116.120401} {\bibfield  {journal}
  {\bibinfo  {journal} {Phys. Rev. Lett.}\ }\textbf {\bibinfo {volume} {116}},\
  \bibinfo {pages} {120401} (\bibinfo {year} {2016})}\BibitemShut {NoStop}%
\bibitem [{\citenamefont {Weidinger}\ and\ \citenamefont
  {Knap}(2017)}]{weidinger2017floquet}%
  \BibitemOpen
  \bibfield  {author} {\bibinfo {author} {\bibfnamefont {S.~A.}\ \bibnamefont
  {Weidinger}}\ and\ \bibinfo {author} {\bibfnamefont {M.}~\bibnamefont
  {Knap}},\ }\bibfield  {title} {\bibinfo {title} {Floquet prethermalization
  and regimes of heating in a periodically driven, interacting quantum
  system},\ }\bibfield  {journal} {\bibinfo  {journal} {Scientific Reports}\
  }\textbf {\bibinfo {volume} {7}},\ \href {https://doi.org/10.1038/srep45382}
  {10.1038/srep45382} (\bibinfo {year} {2017})\BibitemShut {NoStop}%
\bibitem [{\citenamefont {Abanin}\ \emph {et~al.}(2017)\citenamefont {Abanin},
  \citenamefont {De~Roeck}, \citenamefont {Ho},\ and\ \citenamefont
  {Huveneers}}]{abanin2017rigorous}%
  \BibitemOpen
  \bibfield  {author} {\bibinfo {author} {\bibfnamefont {D.}~\bibnamefont
  {Abanin}}, \bibinfo {author} {\bibfnamefont {W.}~\bibnamefont {De~Roeck}},
  \bibinfo {author} {\bibfnamefont {W.~W.}\ \bibnamefont {Ho}},\ and\ \bibinfo
  {author} {\bibfnamefont {F.}~\bibnamefont {Huveneers}},\ }\bibfield  {title}
  {\bibinfo {title} {A rigorous theory of many-body prethermalization for
  periodically driven and closed quantum systems},\ }\href
  {https://doi.org/10.1007/s00220-017-2930-x} {\bibfield  {journal} {\bibinfo
  {journal} {Communications in Mathematical Physics}\ }\textbf {\bibinfo
  {volume} {354}},\ \bibinfo {pages} {809–827} (\bibinfo {year}
  {2017})}\BibitemShut {NoStop}%
\bibitem [{\citenamefont {Mori}(2018)}]{mori2018floquet}%
  \BibitemOpen
  \bibfield  {author} {\bibinfo {author} {\bibfnamefont {T.}~\bibnamefont
  {Mori}},\ }\bibfield  {title} {\bibinfo {title} {Floquet prethermalization in
  periodically driven classical spin systems},\ }\bibfield  {journal} {\bibinfo
   {journal} {Physical Review B}\ }\textbf {\bibinfo {volume} {98}},\ \href
  {https://doi.org/10.1103/physrevb.98.104303} {10.1103/physrevb.98.104303}
  (\bibinfo {year} {2018})\BibitemShut {NoStop}%
\bibitem [{\citenamefont {Mallayya}\ and\ \citenamefont
  {Rigol}(2019)}]{mallayya2019heating}%
  \BibitemOpen
  \bibfield  {author} {\bibinfo {author} {\bibfnamefont {K.}~\bibnamefont
  {Mallayya}}\ and\ \bibinfo {author} {\bibfnamefont {M.}~\bibnamefont
  {Rigol}},\ }\bibfield  {title} {\bibinfo {title} {Heating rates in
  periodically driven strongly interacting quantum many-body systems},\
  }\bibfield  {journal} {\bibinfo  {journal} {Physical Review Letters}\
  }\textbf {\bibinfo {volume} {123}},\ \href
  {https://doi.org/10.1103/physrevlett.123.240603}
  {10.1103/physrevlett.123.240603} (\bibinfo {year} {2019})\BibitemShut
  {NoStop}%
\bibitem [{\citenamefont {Mallayya}\ \emph {et~al.}(2019)\citenamefont
  {Mallayya}, \citenamefont {Rigol},\ and\ \citenamefont
  {De~Roeck}}]{mallayya2019prethermalization}%
  \BibitemOpen
  \bibfield  {author} {\bibinfo {author} {\bibfnamefont {K.}~\bibnamefont
  {Mallayya}}, \bibinfo {author} {\bibfnamefont {M.}~\bibnamefont {Rigol}},\
  and\ \bibinfo {author} {\bibfnamefont {W.}~\bibnamefont {De~Roeck}},\
  }\bibfield  {title} {\bibinfo {title} {Prethermalization and thermalization
  in isolated quantum systems},\ }\bibfield  {journal} {\bibinfo  {journal}
  {Physical Review X}\ }\textbf {\bibinfo {volume} {9}},\ \href
  {https://doi.org/10.1103/physrevx.9.021027} {10.1103/physrevx.9.021027}
  (\bibinfo {year} {2019})\BibitemShut {NoStop}%
\bibitem [{\citenamefont {Kuhlenkamp}\ and\ \citenamefont
  {Knap}(2020)}]{kuhlenkamp2020periodically}%
  \BibitemOpen
  \bibfield  {author} {\bibinfo {author} {\bibfnamefont {C.}~\bibnamefont
  {Kuhlenkamp}}\ and\ \bibinfo {author} {\bibfnamefont {M.}~\bibnamefont
  {Knap}},\ }\bibfield  {title} {\bibinfo {title} {Periodically driven
  sachdev-ye-kitaev models},\ }\bibfield  {journal} {\bibinfo  {journal}
  {Physical Review Letters}\ }\textbf {\bibinfo {volume} {124}},\ \href
  {https://doi.org/10.1103/physrevlett.124.106401}
  {10.1103/physrevlett.124.106401} (\bibinfo {year} {2020})\BibitemShut
  {NoStop}%
\bibitem [{\citenamefont {Fleckenstein}\ and\ \citenamefont
  {Bukov}(2021{\natexlab{a}})}]{fleckenstein2021prethermalization}%
  \BibitemOpen
  \bibfield  {author} {\bibinfo {author} {\bibfnamefont {C.}~\bibnamefont
  {Fleckenstein}}\ and\ \bibinfo {author} {\bibfnamefont {M.}~\bibnamefont
  {Bukov}},\ }\bibfield  {title} {\bibinfo {title} {Prethermalization and
  thermalization in periodically driven many-body systems away from the
  high-frequency limit},\ }\bibfield  {journal} {\bibinfo  {journal} {Physical
  Review B}\ }\textbf {\bibinfo {volume} {103}},\ \href
  {https://doi.org/10.1103/physrevb.103.l140302} {10.1103/physrevb.103.l140302}
  (\bibinfo {year} {2021}{\natexlab{a}})\BibitemShut {NoStop}%
\bibitem [{\citenamefont {Fleckenstein}\ and\ \citenamefont
  {Bukov}(2021{\natexlab{b}})}]{fleckenstein2021thermalization}%
  \BibitemOpen
  \bibfield  {author} {\bibinfo {author} {\bibfnamefont {C.}~\bibnamefont
  {Fleckenstein}}\ and\ \bibinfo {author} {\bibfnamefont {M.}~\bibnamefont
  {Bukov}},\ }\bibfield  {title} {\bibinfo {title} {Thermalization and
  prethermalization in periodically kicked quantum spin chains},\ }\bibfield
  {journal} {\bibinfo  {journal} {Physical Review B}\ }\textbf {\bibinfo
  {volume} {103}},\ \href {https://doi.org/10.1103/physrevb.103.144307}
  {10.1103/physrevb.103.144307} (\bibinfo {year}
  {2021}{\natexlab{b}})\BibitemShut {NoStop}%
\bibitem [{\citenamefont {O’Dea}\ \emph {et~al.}(2024)\citenamefont
  {O’Dea}, \citenamefont {Burnell}, \citenamefont {Chandran},\ and\
  \citenamefont {Khemani}}]{oDea2024prethermal}%
  \BibitemOpen
  \bibfield  {author} {\bibinfo {author} {\bibfnamefont {N.}~\bibnamefont
  {O’Dea}}, \bibinfo {author} {\bibfnamefont {F.}~\bibnamefont {Burnell}},
  \bibinfo {author} {\bibfnamefont {A.}~\bibnamefont {Chandran}},\ and\
  \bibinfo {author} {\bibfnamefont {V.}~\bibnamefont {Khemani}},\ }\bibfield
  {title} {\bibinfo {title} {Prethermal stability of eigenstates under high
  frequency floquet driving},\ }\bibfield  {journal} {\bibinfo  {journal}
  {Physical Review Letters}\ }\textbf {\bibinfo {volume} {132}},\ \href
  {https://doi.org/10.1103/physrevlett.132.100401}
  {10.1103/physrevlett.132.100401} (\bibinfo {year} {2024})\BibitemShut
  {NoStop}%
\bibitem [{\citenamefont {Zhao}\ \emph {et~al.}(2021)\citenamefont {Zhao},
  \citenamefont {Mintert}, \citenamefont {Moessner},\ and\ \citenamefont
  {Knolle}}]{zhao2021random}%
  \BibitemOpen
  \bibfield  {author} {\bibinfo {author} {\bibfnamefont {H.}~\bibnamefont
  {Zhao}}, \bibinfo {author} {\bibfnamefont {F.}~\bibnamefont {Mintert}},
  \bibinfo {author} {\bibfnamefont {R.}~\bibnamefont {Moessner}},\ and\
  \bibinfo {author} {\bibfnamefont {J.}~\bibnamefont {Knolle}},\ }\bibfield
  {title} {\bibinfo {title} {Random multipolar driving: Tunably slow heating
  through spectral engineering},\ }\bibfield  {journal} {\bibinfo  {journal}
  {Physical Review Letters}\ }\textbf {\bibinfo {volume} {126}},\ \href
  {https://doi.org/10.1103/physrevlett.126.040601}
  {10.1103/physrevlett.126.040601} (\bibinfo {year} {2021})\BibitemShut
  {NoStop}%
\bibitem [{\citenamefont {Pawłowski}\ \emph {et~al.}(2024)\citenamefont
  {Pawłowski}, \citenamefont {Panfil}, \citenamefont {Herbrych},\ and\
  \citenamefont {Mierzejewski}}]{pawowski2024long}%
  \BibitemOpen
  \bibfield  {author} {\bibinfo {author} {\bibfnamefont {J.}~\bibnamefont
  {Pawłowski}}, \bibinfo {author} {\bibfnamefont {M.}~\bibnamefont {Panfil}},
  \bibinfo {author} {\bibfnamefont {J.}~\bibnamefont {Herbrych}},\ and\
  \bibinfo {author} {\bibfnamefont {M.}~\bibnamefont {Mierzejewski}},\
  }\bibfield  {title} {\bibinfo {title} {Long-living prethermalization in
  nearly integrable spin ladders},\ }\bibfield  {journal} {\bibinfo  {journal}
  {Physical Review B}\ }\textbf {\bibinfo {volume} {109}},\ \href
  {https://doi.org/10.1103/physrevb.109.l161109} {10.1103/physrevb.109.l161109}
  (\bibinfo {year} {2024})\BibitemShut {NoStop}%
\bibitem [{\citenamefont {McRoberts}\ \emph {et~al.}(2023)\citenamefont
  {McRoberts}, \citenamefont {Zhao}, \citenamefont {Moessner},\ and\
  \citenamefont {Bukov}}]{mcRoberts2023prethermalization}%
  \BibitemOpen
  \bibfield  {author} {\bibinfo {author} {\bibfnamefont {A.~J.}\ \bibnamefont
  {McRoberts}}, \bibinfo {author} {\bibfnamefont {H.}~\bibnamefont {Zhao}},
  \bibinfo {author} {\bibfnamefont {R.}~\bibnamefont {Moessner}},\ and\
  \bibinfo {author} {\bibfnamefont {M.}~\bibnamefont {Bukov}},\ }\bibfield
  {title} {\bibinfo {title} {Prethermalization in periodically driven
  nonreciprocal many-body spin systems},\ }\bibfield  {journal} {\bibinfo
  {journal} {Physical Review Research}\ }\textbf {\bibinfo {volume} {5}},\
  \href {https://doi.org/10.1103/physrevresearch.5.043008}
  {10.1103/physrevresearch.5.043008} (\bibinfo {year} {2023})\BibitemShut
  {NoStop}%
\bibitem [{\citenamefont {Jin}\ \emph {et~al.}(2023)\citenamefont {Jin},
  \citenamefont {Knolle},\ and\ \citenamefont {Knap}}]{jin2023fractionalized}%
  \BibitemOpen
  \bibfield  {author} {\bibinfo {author} {\bibfnamefont {H.-K.}\ \bibnamefont
  {Jin}}, \bibinfo {author} {\bibfnamefont {J.}~\bibnamefont {Knolle}},\ and\
  \bibinfo {author} {\bibfnamefont {M.}~\bibnamefont {Knap}},\ }\bibfield
  {title} {\bibinfo {title} {Fractionalized prethermalization in a driven
  quantum spin liquid},\ }\bibfield  {journal} {\bibinfo  {journal} {Physical
  Review Letters}\ }\textbf {\bibinfo {volume} {130}},\ \href
  {https://doi.org/10.1103/physrevlett.130.226701}
  {10.1103/physrevlett.130.226701} (\bibinfo {year} {2023})\BibitemShut
  {NoStop}%
\bibitem [{\citenamefont {Kitaev}(2006)}]{kitaev2006anyons}%
  \BibitemOpen
  \bibfield  {author} {\bibinfo {author} {\bibfnamefont {A.}~\bibnamefont
  {Kitaev}},\ }\bibfield  {title} {\bibinfo {title} {Anyons in an exactly
  solved model and beyond},\ }\href
  {https://doi.org/https://doi.org/10.1016/j.aop.2005.10.005} {\bibfield
  {journal} {\bibinfo  {journal} {Annals of Physics}\ }\textbf {\bibinfo
  {volume} {321}},\ \bibinfo {pages} {2} (\bibinfo {year} {2006})},\ \bibinfo
  {note} {january Special Issue}\BibitemShut {NoStop}%
\bibitem [{\citenamefont {Meden}\ and\ \citenamefont
  {Sch\"{o}nhammer}(1992)}]{meden1992spectral}%
  \BibitemOpen
  \bibfield  {author} {\bibinfo {author} {\bibfnamefont {V.}~\bibnamefont
  {Meden}}\ and\ \bibinfo {author} {\bibfnamefont {K.}~\bibnamefont
  {Sch\"{o}nhammer}},\ }\bibfield  {title} {\bibinfo {title} {Spectral
  functions for the tomonaga-luttinger model},\ }\href
  {https://doi.org/10.1103/physrevb.46.15753} {\bibfield  {journal} {\bibinfo
  {journal} {Physical Review B}\ }\textbf {\bibinfo {volume} {46}},\ \bibinfo
  {pages} {15753–15760} (\bibinfo {year} {1992})}\BibitemShut {NoStop}%
\bibitem [{\citenamefont {Voit}(1993)}]{voit1993charge}%
  \BibitemOpen
  \bibfield  {author} {\bibinfo {author} {\bibfnamefont {J.}~\bibnamefont
  {Voit}},\ }\bibfield  {title} {\bibinfo {title} {Charge-spin separation and
  the spectral properties of luttinger liquids},\ }\href
  {https://doi.org/10.1088/0953-8984/5/44/020} {\bibfield  {journal} {\bibinfo
  {journal} {Journal of Physics: Condensed Matter}\ }\textbf {\bibinfo {volume}
  {5}},\ \bibinfo {pages} {8305–8336} (\bibinfo {year} {1993})}\BibitemShut
  {NoStop}%
\bibitem [{\citenamefont {Giamarchi}(2003)}]{giamarchi2003quantum}%
  \BibitemOpen
  \bibfield  {author} {\bibinfo {author} {\bibfnamefont {T.}~\bibnamefont
  {Giamarchi}},\ }\href@noop {} {\emph {\bibinfo {title} {Quantum physics in
  one dimension}}},\ Vol.\ \bibinfo {volume} {121}\ (\bibinfo  {publisher}
  {Clarendon press},\ \bibinfo {year} {2003})\BibitemShut {NoStop}%
\bibitem [{\citenamefont {Vijayan}\ \emph {et~al.}(2020)\citenamefont
  {Vijayan}, \citenamefont {Sompet}, \citenamefont {Salomon}, \citenamefont
  {Koepsell}, \citenamefont {Hirthe}, \citenamefont {Bohrdt}, \citenamefont
  {Grusdt}, \citenamefont {Bloch},\ and\ \citenamefont
  {Gross}}]{viJayan2020time}%
  \BibitemOpen
  \bibfield  {author} {\bibinfo {author} {\bibfnamefont {J.}~\bibnamefont
  {Vijayan}}, \bibinfo {author} {\bibfnamefont {P.}~\bibnamefont {Sompet}},
  \bibinfo {author} {\bibfnamefont {G.}~\bibnamefont {Salomon}}, \bibinfo
  {author} {\bibfnamefont {J.}~\bibnamefont {Koepsell}}, \bibinfo {author}
  {\bibfnamefont {S.}~\bibnamefont {Hirthe}}, \bibinfo {author} {\bibfnamefont
  {A.}~\bibnamefont {Bohrdt}}, \bibinfo {author} {\bibfnamefont
  {F.}~\bibnamefont {Grusdt}}, \bibinfo {author} {\bibfnamefont
  {I.}~\bibnamefont {Bloch}},\ and\ \bibinfo {author} {\bibfnamefont
  {C.}~\bibnamefont {Gross}},\ }\bibfield  {title} {\bibinfo {title}
  {Time-resolved observation of spin-charge deconfinement in fermionic hubbard
  chains},\ }\href {https://doi.org/10.1126/science.aay2354} {\bibfield
  {journal} {\bibinfo  {journal} {Science}\ }\textbf {\bibinfo {volume}
  {367}},\ \bibinfo {pages} {186–189} (\bibinfo {year} {2020})}\BibitemShut
  {NoStop}%
\bibitem [{\citenamefont {Aidelsburger}\ \emph {et~al.}(2013)\citenamefont
  {Aidelsburger}, \citenamefont {Atala}, \citenamefont {Lohse}, \citenamefont
  {Barreiro}, \citenamefont {Paredes},\ and\ \citenamefont
  {Bloch}}]{aidelsburger2013realization}%
  \BibitemOpen
  \bibfield  {author} {\bibinfo {author} {\bibfnamefont {M.}~\bibnamefont
  {Aidelsburger}}, \bibinfo {author} {\bibfnamefont {M.}~\bibnamefont {Atala}},
  \bibinfo {author} {\bibfnamefont {M.}~\bibnamefont {Lohse}}, \bibinfo
  {author} {\bibfnamefont {J.~T.}\ \bibnamefont {Barreiro}}, \bibinfo {author}
  {\bibfnamefont {B.}~\bibnamefont {Paredes}},\ and\ \bibinfo {author}
  {\bibfnamefont {I.}~\bibnamefont {Bloch}},\ }\bibfield  {title} {\bibinfo
  {title} {Realization of the hofstadter hamiltonian with ultracold atoms in
  optical lattices},\ }\bibfield  {journal} {\bibinfo  {journal} {Physical
  Review Letters}\ }\textbf {\bibinfo {volume} {111}},\ \href
  {https://doi.org/10.1103/physrevlett.111.185301}
  {10.1103/physrevlett.111.185301} (\bibinfo {year} {2013})\BibitemShut
  {NoStop}%
\bibitem [{\citenamefont {Miyake}\ \emph {et~al.}(2013)\citenamefont {Miyake},
  \citenamefont {Siviloglou}, \citenamefont {Kennedy}, \citenamefont {Burton},\
  and\ \citenamefont {Ketterle}}]{miyake2013realizing}%
  \BibitemOpen
  \bibfield  {author} {\bibinfo {author} {\bibfnamefont {H.}~\bibnamefont
  {Miyake}}, \bibinfo {author} {\bibfnamefont {G.~A.}\ \bibnamefont
  {Siviloglou}}, \bibinfo {author} {\bibfnamefont {C.~J.}\ \bibnamefont
  {Kennedy}}, \bibinfo {author} {\bibfnamefont {W.~C.}\ \bibnamefont
  {Burton}},\ and\ \bibinfo {author} {\bibfnamefont {W.}~\bibnamefont
  {Ketterle}},\ }\bibfield  {title} {\bibinfo {title} {Realizing the harper
  hamiltonian with laser-assisted tunneling in optical lattices},\ }\bibfield
  {journal} {\bibinfo  {journal} {Physical Review Letters}\ }\textbf {\bibinfo
  {volume} {111}},\ \href {https://doi.org/10.1103/physrevlett.111.185302}
  {10.1103/physrevlett.111.185302} (\bibinfo {year} {2013})\BibitemShut
  {NoStop}%
\bibitem [{\citenamefont {Jotzu}\ \emph {et~al.}(2014)\citenamefont {Jotzu},
  \citenamefont {Messer}, \citenamefont {Desbuquois}, \citenamefont {Lebrat},
  \citenamefont {Uehlinger}, \citenamefont {Greif},\ and\ \citenamefont
  {Esslinger}}]{jotzu2014experimental}%
  \BibitemOpen
  \bibfield  {author} {\bibinfo {author} {\bibfnamefont {G.}~\bibnamefont
  {Jotzu}}, \bibinfo {author} {\bibfnamefont {M.}~\bibnamefont {Messer}},
  \bibinfo {author} {\bibfnamefont {R.}~\bibnamefont {Desbuquois}}, \bibinfo
  {author} {\bibfnamefont {M.}~\bibnamefont {Lebrat}}, \bibinfo {author}
  {\bibfnamefont {T.}~\bibnamefont {Uehlinger}}, \bibinfo {author}
  {\bibfnamefont {D.}~\bibnamefont {Greif}},\ and\ \bibinfo {author}
  {\bibfnamefont {T.}~\bibnamefont {Esslinger}},\ }\bibfield  {title} {\bibinfo
  {title} {Experimental realization of the topological haldane model with
  ultracold fermions},\ }\href {https://doi.org/10.1038/nature13915} {\bibfield
   {journal} {\bibinfo  {journal} {Nature}\ }\textbf {\bibinfo {volume}
  {515}},\ \bibinfo {pages} {237} (\bibinfo {year} {2014})}\BibitemShut
  {NoStop}%
\bibitem [{\citenamefont {Bordia}\ \emph {et~al.}(2017)\citenamefont {Bordia},
  \citenamefont {L{\"u}schen}, \citenamefont {Schneider}, \citenamefont
  {Knap},\ and\ \citenamefont {Bloch}}]{bordia2017periodically}%
  \BibitemOpen
  \bibfield  {author} {\bibinfo {author} {\bibfnamefont {P.}~\bibnamefont
  {Bordia}}, \bibinfo {author} {\bibfnamefont {H.}~\bibnamefont {L{\"u}schen}},
  \bibinfo {author} {\bibfnamefont {U.}~\bibnamefont {Schneider}}, \bibinfo
  {author} {\bibfnamefont {M.}~\bibnamefont {Knap}},\ and\ \bibinfo {author}
  {\bibfnamefont {I.}~\bibnamefont {Bloch}},\ }\bibfield  {title} {\bibinfo
  {title} {Periodically driving a many-body localized quantum system},\ }\href
  {https://doi.org/https://doi.org/10.1038/nphys4020} {\bibfield  {journal}
  {\bibinfo  {journal} {Nature Physics}\ }\textbf {\bibinfo {volume} {13}},\
  \bibinfo {pages} {460} (\bibinfo {year} {2017})}\BibitemShut {NoStop}%
\bibitem [{\citenamefont {Rubio-Abadal}\ \emph {et~al.}(2020)\citenamefont
  {Rubio-Abadal}, \citenamefont {Ippoliti}, \citenamefont {Hollerith},
  \citenamefont {Wei}, \citenamefont {Rui}, \citenamefont {Sondhi},
  \citenamefont {Khemani}, \citenamefont {Gross},\ and\ \citenamefont
  {Bloch}}]{rubioAbadal2020floquet}%
  \BibitemOpen
  \bibfield  {author} {\bibinfo {author} {\bibfnamefont {A.}~\bibnamefont
  {Rubio-Abadal}}, \bibinfo {author} {\bibfnamefont {M.}~\bibnamefont
  {Ippoliti}}, \bibinfo {author} {\bibfnamefont {S.}~\bibnamefont {Hollerith}},
  \bibinfo {author} {\bibfnamefont {D.}~\bibnamefont {Wei}}, \bibinfo {author}
  {\bibfnamefont {J.}~\bibnamefont {Rui}}, \bibinfo {author} {\bibfnamefont
  {S.}~\bibnamefont {Sondhi}}, \bibinfo {author} {\bibfnamefont
  {V.}~\bibnamefont {Khemani}}, \bibinfo {author} {\bibfnamefont
  {C.}~\bibnamefont {Gross}},\ and\ \bibinfo {author} {\bibfnamefont
  {I.}~\bibnamefont {Bloch}},\ }\bibfield  {title} {\bibinfo {title} {Floquet
  prethermalization in a bose-hubbard system},\ }\bibfield  {journal} {\bibinfo
   {journal} {Physical Review X}\ }\textbf {\bibinfo {volume} {10}},\ \href
  {https://doi.org/10.1103/physrevx.10.021044} {10.1103/physrevx.10.021044}
  (\bibinfo {year} {2020})\BibitemShut {NoStop}%
\bibitem [{\citenamefont {Auerbach}(1994)}]{auerbach1994interacting}%
  \BibitemOpen
  \bibfield  {author} {\bibinfo {author} {\bibfnamefont {A.}~\bibnamefont
  {Auerbach}},\ }\href {https://doi.org/10.1007/978-1-4612-0869-3} {\emph
  {\bibinfo {title} {Interacting Electrons and Quantum Magnetism}}}\ (\bibinfo
  {publisher} {Springer New York},\ \bibinfo {year} {1994})\BibitemShut
  {NoStop}%
\bibitem [{\citenamefont {Bohrdt}\ \emph {et~al.}(2018)\citenamefont {Bohrdt},
  \citenamefont {Greif}, \citenamefont {Demler}, \citenamefont {Knap},\ and\
  \citenamefont {Grusdt}}]{bohrdt2018angle}%
  \BibitemOpen
  \bibfield  {author} {\bibinfo {author} {\bibfnamefont {A.}~\bibnamefont
  {Bohrdt}}, \bibinfo {author} {\bibfnamefont {D.}~\bibnamefont {Greif}},
  \bibinfo {author} {\bibfnamefont {E.}~\bibnamefont {Demler}}, \bibinfo
  {author} {\bibfnamefont {M.}~\bibnamefont {Knap}},\ and\ \bibinfo {author}
  {\bibfnamefont {F.}~\bibnamefont {Grusdt}},\ }\bibfield  {title} {\bibinfo
  {title} {Angle-resolved photoemission spectroscopy with quantum gas
  microscopes},\ }\bibfield  {journal} {\bibinfo  {journal} {Physical Review
  B}\ }\textbf {\bibinfo {volume} {97}},\ \href
  {https://doi.org/10.1103/physrevb.97.125117} {10.1103/physrevb.97.125117}
  (\bibinfo {year} {2018})\BibitemShut {NoStop}%
\bibitem [{\citenamefont {Bannister}\ and\ \citenamefont
  {d’Ambrumenil}(2000)}]{bannister2000spectral}%
  \BibitemOpen
  \bibfield  {author} {\bibinfo {author} {\bibfnamefont {R.~N.}\ \bibnamefont
  {Bannister}}\ and\ \bibinfo {author} {\bibfnamefont {N.}~\bibnamefont
  {d’Ambrumenil}},\ }\bibfield  {title} {\bibinfo {title} {Spectral functions
  of half-filled one-dimensional hubbard rings with varying boundary
  conditions},\ }\href {https://doi.org/10.1103/physrevb.61.4651} {\bibfield
  {journal} {\bibinfo  {journal} {Physical Review B}\ }\textbf {\bibinfo
  {volume} {61}},\ \bibinfo {pages} {4651–4658} (\bibinfo {year}
  {2000})}\BibitemShut {NoStop}%
\bibitem [{\citenamefont {Bartsch}\ and\ \citenamefont
  {Gemmer}(2009)}]{bartsch2009dynamical}%
  \BibitemOpen
  \bibfield  {author} {\bibinfo {author} {\bibfnamefont {C.}~\bibnamefont
  {Bartsch}}\ and\ \bibinfo {author} {\bibfnamefont {J.}~\bibnamefont
  {Gemmer}},\ }\bibfield  {title} {\bibinfo {title} {Dynamical typicality of
  quantum expectation values},\ }\bibfield  {journal} {\bibinfo  {journal}
  {Physical Review Letters}\ }\textbf {\bibinfo {volume} {102}},\ \href
  {https://doi.org/10.1103/physrevlett.102.110403}
  {10.1103/physrevlett.102.110403} (\bibinfo {year} {2009})\BibitemShut
  {NoStop}%
\bibitem [{\citenamefont {Reimann}(2018)}]{reimann2018dynamical}%
  \BibitemOpen
  \bibfield  {author} {\bibinfo {author} {\bibfnamefont {P.}~\bibnamefont
  {Reimann}},\ }\bibfield  {title} {\bibinfo {title} {Dynamical typicality of
  isolated many-body quantum systems},\ }\bibfield  {journal} {\bibinfo
  {journal} {Physical Review E}\ }\textbf {\bibinfo {volume} {97}},\ \href
  {https://doi.org/10.1103/physreve.97.062129} {10.1103/physreve.97.062129}
  (\bibinfo {year} {2018})\BibitemShut {NoStop}%
\bibitem [{\citenamefont {Sugiura}\ and\ \citenamefont
  {Shimizu}(2012)}]{sugiura2012thermal}%
  \BibitemOpen
  \bibfield  {author} {\bibinfo {author} {\bibfnamefont {S.}~\bibnamefont
  {Sugiura}}\ and\ \bibinfo {author} {\bibfnamefont {A.}~\bibnamefont
  {Shimizu}},\ }\bibfield  {title} {\bibinfo {title} {Thermal pure quantum
  states at finite temperature},\ }\bibfield  {journal} {\bibinfo  {journal}
  {Physical Review Letters}\ }\textbf {\bibinfo {volume} {108}},\ \href
  {https://doi.org/10.1103/physrevlett.108.240401}
  {10.1103/physrevlett.108.240401} (\bibinfo {year} {2012})\BibitemShut
  {NoStop}%
\bibitem [{\citenamefont {Weinberg}\ and\ \citenamefont
  {Bukov}(2019)}]{weinberg2019quspin}%
  \BibitemOpen
  \bibfield  {author} {\bibinfo {author} {\bibfnamefont {P.}~\bibnamefont
  {Weinberg}}\ and\ \bibinfo {author} {\bibfnamefont {M.}~\bibnamefont
  {Bukov}},\ }\bibfield  {title} {\bibinfo {title} {Quspin: a python package
  for dynamics and exact diagonalisation of quantum many body systems. part ii:
  bosons, fermions and higher spins},\ }\bibfield  {journal} {\bibinfo
  {journal} {SciPost Physics}\ }\textbf {\bibinfo {volume} {7}},\ \href
  {https://doi.org/10.21468/scipostphys.7.2.020} {10.21468/scipostphys.7.2.020}
  (\bibinfo {year} {2019})\BibitemShut {NoStop}%
\bibitem [{\citenamefont {Sieberer}\ \emph {et~al.}(2018)\citenamefont
  {Sieberer}, \citenamefont {Rieder}, \citenamefont {Fischer},\ and\
  \citenamefont {Fulga}}]{sieberer2018statistical}%
  \BibitemOpen
  \bibfield  {author} {\bibinfo {author} {\bibfnamefont {L.~M.}\ \bibnamefont
  {Sieberer}}, \bibinfo {author} {\bibfnamefont {M.-T.}\ \bibnamefont
  {Rieder}}, \bibinfo {author} {\bibfnamefont {M.~H.}\ \bibnamefont
  {Fischer}},\ and\ \bibinfo {author} {\bibfnamefont {I.~C.}\ \bibnamefont
  {Fulga}},\ }\bibfield  {title} {\bibinfo {title} {Statistical periodicity in
  driven quantum systems: General formalism and application to noisy floquet
  topological chains},\ }\bibfield  {journal} {\bibinfo  {journal} {Physical
  Review B}\ }\textbf {\bibinfo {volume} {98}},\ \href
  {https://doi.org/10.1103/physrevb.98.214301} {10.1103/physrevb.98.214301}
  (\bibinfo {year} {2018})\BibitemShut {NoStop}%
\bibitem [{\citenamefont {Bukov}\ \emph {et~al.}(2016)\citenamefont {Bukov},
  \citenamefont {Kolodrubetz},\ and\ \citenamefont
  {Polkovnikov}}]{bukov2016schrieffer}%
  \BibitemOpen
  \bibfield  {author} {\bibinfo {author} {\bibfnamefont {M.}~\bibnamefont
  {Bukov}}, \bibinfo {author} {\bibfnamefont {M.}~\bibnamefont {Kolodrubetz}},\
  and\ \bibinfo {author} {\bibfnamefont {A.}~\bibnamefont {Polkovnikov}},\
  }\bibfield  {title} {\bibinfo {title} {Schrieffer-wolff transformation for
  periodically driven systems: Strongly correlated systems with artificial
  gauge fields},\ }\bibfield  {journal} {\bibinfo  {journal} {Physical Review
  Letters}\ }\textbf {\bibinfo {volume} {116}},\ \href
  {https://doi.org/10.1103/physrevlett.116.125301}
  {10.1103/physrevlett.116.125301} (\bibinfo {year} {2016})\BibitemShut
  {NoStop}%
\bibitem [{\citenamefont {De~Roeck}\ and\ \citenamefont
  {Verreet}(2019)}]{deRoeck2019very}%
  \BibitemOpen
  \bibfield  {author} {\bibinfo {author} {\bibfnamefont {W.}~\bibnamefont
  {De~Roeck}}\ and\ \bibinfo {author} {\bibfnamefont {V.}~\bibnamefont
  {Verreet}},\ }\bibfield  {title} {\bibinfo {title} {Very slow heating for
  weakly driven quantum many-body systems},\ }\bibfield  {journal} {\bibinfo
  {journal} {arXiv preprint}\ }\href
  {https://doi.org/10.48550/arXiv.1911.01998} {10.48550/arXiv.1911.01998}
  (\bibinfo {year} {2019})\BibitemShut {NoStop}%
\bibitem [{\citenamefont {Bravyi}\ \emph {et~al.}(2011)\citenamefont {Bravyi},
  \citenamefont {DiVincenzo},\ and\ \citenamefont
  {Loss}}]{bravyi2011schrieffer}%
  \BibitemOpen
  \bibfield  {author} {\bibinfo {author} {\bibfnamefont {S.}~\bibnamefont
  {Bravyi}}, \bibinfo {author} {\bibfnamefont {D.~P.}\ \bibnamefont
  {DiVincenzo}},\ and\ \bibinfo {author} {\bibfnamefont {D.}~\bibnamefont
  {Loss}},\ }\bibfield  {title} {\bibinfo {title} {Schrieffer–wolff
  transformation for quantum many-body systems},\ }\href
  {https://doi.org/10.1016/j.aop.2011.06.004} {\bibfield  {journal} {\bibinfo
  {journal} {Annals of Physics}\ }\textbf {\bibinfo {volume} {326}},\ \bibinfo
  {pages} {2793–2826} (\bibinfo {year} {2011})}\BibitemShut {NoStop}%
\bibitem [{\citenamefont {Wilcox}(1967)}]{wilcox1967exponential}%
  \BibitemOpen
  \bibfield  {author} {\bibinfo {author} {\bibfnamefont {R.~M.}\ \bibnamefont
  {Wilcox}},\ }\bibfield  {title} {\bibinfo {title} {Exponential operators and
  parameter differentiation in quantum physics},\ }\href
  {https://doi.org/10.1063/1.1705306} {\bibfield  {journal} {\bibinfo
  {journal} {Journal of Mathematical Physics}\ }\textbf {\bibinfo {volume}
  {8}},\ \bibinfo {pages} {962–982} (\bibinfo {year} {1967})}\BibitemShut
  {NoStop}%
\bibitem [{\citenamefont {Sun}\ and\ \citenamefont
  {Eckardt}(2020)}]{sun2020optimal}%
  \BibitemOpen
  \bibfield  {author} {\bibinfo {author} {\bibfnamefont {G.}~\bibnamefont
  {Sun}}\ and\ \bibinfo {author} {\bibfnamefont {A.}~\bibnamefont {Eckardt}},\
  }\bibfield  {title} {\bibinfo {title} {Optimal frequency window for floquet
  engineering in optical lattices},\ }\bibfield  {journal} {\bibinfo  {journal}
  {Physical Review Research}\ }\textbf {\bibinfo {volume} {2}},\ \href
  {https://doi.org/10.1103/physrevresearch.2.013241}
  {10.1103/physrevresearch.2.013241} (\bibinfo {year} {2020})\BibitemShut
  {NoStop}%
\bibitem [{\citenamefont {Bloch}\ \emph {et~al.}(2008)\citenamefont {Bloch},
  \citenamefont {Dalibard},\ and\ \citenamefont {Zwerger}}]{bloch2008many}%
  \BibitemOpen
  \bibfield  {author} {\bibinfo {author} {\bibfnamefont {I.}~\bibnamefont
  {Bloch}}, \bibinfo {author} {\bibfnamefont {J.}~\bibnamefont {Dalibard}},\
  and\ \bibinfo {author} {\bibfnamefont {W.}~\bibnamefont {Zwerger}},\
  }\bibfield  {title} {\bibinfo {title} {Many-body physics with ultracold
  gases},\ }\href {https://doi.org/10.1103/revmodphys.80.885} {\bibfield
  {journal} {\bibinfo  {journal} {Reviews of Modern Physics}\ }\textbf
  {\bibinfo {volume} {80}},\ \bibinfo {pages} {885–964} (\bibinfo {year}
  {2008})}\BibitemShut {NoStop}%
\bibitem [{\citenamefont {Wei}\ \emph {et~al.}(2022)\citenamefont {Wei},
  \citenamefont {Rubio-Abadal}, \citenamefont {Ye}, \citenamefont {Machado},
  \citenamefont {Kemp}, \citenamefont {Srakaew}, \citenamefont {Hollerith},
  \citenamefont {Rui}, \citenamefont {Gopalakrishnan}, \citenamefont {Yao},
  \citenamefont {Bloch},\ and\ \citenamefont {Zeiher}}]{wei2022quantum}%
  \BibitemOpen
  \bibfield  {author} {\bibinfo {author} {\bibfnamefont {D.}~\bibnamefont
  {Wei}}, \bibinfo {author} {\bibfnamefont {A.}~\bibnamefont {Rubio-Abadal}},
  \bibinfo {author} {\bibfnamefont {B.}~\bibnamefont {Ye}}, \bibinfo {author}
  {\bibfnamefont {F.}~\bibnamefont {Machado}}, \bibinfo {author} {\bibfnamefont
  {J.}~\bibnamefont {Kemp}}, \bibinfo {author} {\bibfnamefont {K.}~\bibnamefont
  {Srakaew}}, \bibinfo {author} {\bibfnamefont {S.}~\bibnamefont {Hollerith}},
  \bibinfo {author} {\bibfnamefont {J.}~\bibnamefont {Rui}}, \bibinfo {author}
  {\bibfnamefont {S.}~\bibnamefont {Gopalakrishnan}}, \bibinfo {author}
  {\bibfnamefont {N.~Y.}\ \bibnamefont {Yao}}, \bibinfo {author} {\bibfnamefont
  {I.}~\bibnamefont {Bloch}},\ and\ \bibinfo {author} {\bibfnamefont
  {J.}~\bibnamefont {Zeiher}},\ }\bibfield  {title} {\bibinfo {title} {Quantum
  gas microscopy of kardar-parisi-zhang superdiffusion},\ }\href
  {https://doi.org/10.1126/science.abk2397} {\bibfield  {journal} {\bibinfo
  {journal} {Science}\ }\textbf {\bibinfo {volume} {376}},\ \bibinfo {pages}
  {716–720} (\bibinfo {year} {2022})}\BibitemShut {NoStop}%
\bibitem [{\citenamefont {Wybo}\ \emph {et~al.}(2023)\citenamefont {Wybo},
  \citenamefont {Bastianello}, \citenamefont {Aidelsburger}, \citenamefont
  {Bloch},\ and\ \citenamefont {Knap}}]{wybo2023preparing}%
  \BibitemOpen
  \bibfield  {author} {\bibinfo {author} {\bibfnamefont {E.}~\bibnamefont
  {Wybo}}, \bibinfo {author} {\bibfnamefont {A.}~\bibnamefont {Bastianello}},
  \bibinfo {author} {\bibfnamefont {M.}~\bibnamefont {Aidelsburger}}, \bibinfo
  {author} {\bibfnamefont {I.}~\bibnamefont {Bloch}},\ and\ \bibinfo {author}
  {\bibfnamefont {M.}~\bibnamefont {Knap}},\ }\bibfield  {title} {\bibinfo
  {title} {Preparing and analyzing solitons in the sine-gordon model with
  quantum gas microscopes},\ }\bibfield  {journal} {\bibinfo  {journal} {PRX
  Quantum}\ }\textbf {\bibinfo {volume} {4}},\ \href
  {https://doi.org/10.1103/prxquantum.4.030308} {10.1103/prxquantum.4.030308}
  (\bibinfo {year} {2023})\BibitemShut {NoStop}%
\bibitem [{\citenamefont {Impertro}\ \emph {et~al.}(2024)\citenamefont
  {Impertro}, \citenamefont {Karch}, \citenamefont {Wienand}, \citenamefont
  {Huh}, \citenamefont {Schweizer}, \citenamefont {Bloch},\ and\ \citenamefont
  {Aidelsburger}}]{impertro2023local}%
  \BibitemOpen
  \bibfield  {author} {\bibinfo {author} {\bibfnamefont {A.}~\bibnamefont
  {Impertro}}, \bibinfo {author} {\bibfnamefont {S.}~\bibnamefont {Karch}},
  \bibinfo {author} {\bibfnamefont {J.~F.}\ \bibnamefont {Wienand}}, \bibinfo
  {author} {\bibfnamefont {S.}~\bibnamefont {Huh}}, \bibinfo {author}
  {\bibfnamefont {C.}~\bibnamefont {Schweizer}}, \bibinfo {author}
  {\bibfnamefont {I.}~\bibnamefont {Bloch}},\ and\ \bibinfo {author}
  {\bibfnamefont {M.}~\bibnamefont {Aidelsburger}},\ }\bibfield  {title}
  {\bibinfo {title} {Local readout and control of current and kinetic energy
  operators in optical lattices},\ }\bibfield  {journal} {\bibinfo  {journal}
  {Physical Review Letters}\ }\textbf {\bibinfo {volume} {133}},\ \href
  {https://doi.org/10.1103/physrevlett.133.063401}
  {10.1103/physrevlett.133.063401} (\bibinfo {year} {2024})\BibitemShut
  {NoStop}%
\bibitem [{\citenamefont {Else}\ \emph {et~al.}(2020)\citenamefont {Else},
  \citenamefont {Monroe}, \citenamefont {Nayak},\ and\ \citenamefont
  {Yao}}]{else2020discrete}%
  \BibitemOpen
  \bibfield  {author} {\bibinfo {author} {\bibfnamefont {D.~V.}\ \bibnamefont
  {Else}}, \bibinfo {author} {\bibfnamefont {C.}~\bibnamefont {Monroe}},
  \bibinfo {author} {\bibfnamefont {C.}~\bibnamefont {Nayak}},\ and\ \bibinfo
  {author} {\bibfnamefont {N.~Y.}\ \bibnamefont {Yao}},\ }\bibfield  {title}
  {\bibinfo {title} {Discrete time crystals},\ }\href
  {https://doi.org/10.1146/annurev-conmatphys-031119-050658} {\bibfield
  {journal} {\bibinfo  {journal} {Annual Review of Condensed Matter Physics}\
  }\textbf {\bibinfo {volume} {11}},\ \bibinfo {pages} {467–499} (\bibinfo
  {year} {2020})}\BibitemShut {NoStop}%
\bibitem [{\citenamefont {Zhang}\ \emph {et~al.}(2017)\citenamefont {Zhang},
  \citenamefont {Hess}, \citenamefont {Kyprianidis}, \citenamefont {Becker},
  \citenamefont {Lee}, \citenamefont {Smith}, \citenamefont {Pagano},
  \citenamefont {Potirniche}, \citenamefont {Potter}, \citenamefont
  {Vishwanath}, \citenamefont {Yao},\ and\ \citenamefont
  {Monroe}}]{zhang2017observation}%
  \BibitemOpen
  \bibfield  {author} {\bibinfo {author} {\bibfnamefont {J.}~\bibnamefont
  {Zhang}}, \bibinfo {author} {\bibfnamefont {P.~W.}\ \bibnamefont {Hess}},
  \bibinfo {author} {\bibfnamefont {A.}~\bibnamefont {Kyprianidis}}, \bibinfo
  {author} {\bibfnamefont {P.}~\bibnamefont {Becker}}, \bibinfo {author}
  {\bibfnamefont {A.}~\bibnamefont {Lee}}, \bibinfo {author} {\bibfnamefont
  {J.}~\bibnamefont {Smith}}, \bibinfo {author} {\bibfnamefont
  {G.}~\bibnamefont {Pagano}}, \bibinfo {author} {\bibfnamefont {I.-D.}\
  \bibnamefont {Potirniche}}, \bibinfo {author} {\bibfnamefont {A.~C.}\
  \bibnamefont {Potter}}, \bibinfo {author} {\bibfnamefont {A.}~\bibnamefont
  {Vishwanath}}, \bibinfo {author} {\bibfnamefont {N.~Y.}\ \bibnamefont
  {Yao}},\ and\ \bibinfo {author} {\bibfnamefont {C.}~\bibnamefont {Monroe}},\
  }\bibfield  {title} {\bibinfo {title} {Observation of a discrete time
  crystal},\ }\href {https://doi.org/10.1038/nature21413} {\bibfield  {journal}
  {\bibinfo  {journal} {Nature}\ }\textbf {\bibinfo {volume} {543}},\ \bibinfo
  {pages} {217–220} (\bibinfo {year} {2017})}\BibitemShut {NoStop}%
\bibitem [{\citenamefont {Zaletel}\ \emph {et~al.}(2023)\citenamefont
  {Zaletel}, \citenamefont {Lukin}, \citenamefont {Monroe}, \citenamefont
  {Nayak}, \citenamefont {Wilczek},\ and\ \citenamefont
  {Yao}}]{zaletel2023colloquium}%
  \BibitemOpen
  \bibfield  {author} {\bibinfo {author} {\bibfnamefont {M.~P.}\ \bibnamefont
  {Zaletel}}, \bibinfo {author} {\bibfnamefont {M.}~\bibnamefont {Lukin}},
  \bibinfo {author} {\bibfnamefont {C.}~\bibnamefont {Monroe}}, \bibinfo
  {author} {\bibfnamefont {C.}~\bibnamefont {Nayak}}, \bibinfo {author}
  {\bibfnamefont {F.}~\bibnamefont {Wilczek}},\ and\ \bibinfo {author}
  {\bibfnamefont {N.~Y.}\ \bibnamefont {Yao}},\ }\bibfield  {title} {\bibinfo
  {title} {Colloquium: Quantum and classical discrete time crystals},\
  }\bibfield  {journal} {\bibinfo  {journal} {Reviews of Modern Physics}\
  }\textbf {\bibinfo {volume} {95}},\ \href
  {https://doi.org/10.1103/revmodphys.95.031001} {10.1103/revmodphys.95.031001}
  (\bibinfo {year} {2023})\BibitemShut {NoStop}%
\bibitem [{\citenamefont {Martinez}\ \emph {et~al.}(2016)\citenamefont
  {Martinez}, \citenamefont {Muschik}, \citenamefont {Schindler}, \citenamefont
  {Nigg}, \citenamefont {Erhard}, \citenamefont {Heyl}, \citenamefont {Hauke},
  \citenamefont {Dalmonte}, \citenamefont {Monz}, \citenamefont {Zoller},\ and\
  \citenamefont {Blatt}}]{martinez2016real}%
  \BibitemOpen
  \bibfield  {author} {\bibinfo {author} {\bibfnamefont {E.~A.}\ \bibnamefont
  {Martinez}}, \bibinfo {author} {\bibfnamefont {C.~A.}\ \bibnamefont
  {Muschik}}, \bibinfo {author} {\bibfnamefont {P.}~\bibnamefont {Schindler}},
  \bibinfo {author} {\bibfnamefont {D.}~\bibnamefont {Nigg}}, \bibinfo {author}
  {\bibfnamefont {A.}~\bibnamefont {Erhard}}, \bibinfo {author} {\bibfnamefont
  {M.}~\bibnamefont {Heyl}}, \bibinfo {author} {\bibfnamefont {P.}~\bibnamefont
  {Hauke}}, \bibinfo {author} {\bibfnamefont {M.}~\bibnamefont {Dalmonte}},
  \bibinfo {author} {\bibfnamefont {T.}~\bibnamefont {Monz}}, \bibinfo {author}
  {\bibfnamefont {P.}~\bibnamefont {Zoller}},\ and\ \bibinfo {author}
  {\bibfnamefont {R.}~\bibnamefont {Blatt}},\ }\bibfield  {title} {\bibinfo
  {title} {Real-time dynamics of lattice gauge theories with a few-qubit
  quantum computer},\ }\href {https://doi.org/10.1038/nature18318} {\bibfield
  {journal} {\bibinfo  {journal} {Nature}\ }\textbf {\bibinfo {volume} {534}},\
  \bibinfo {pages} {516–519} (\bibinfo {year} {2016})}\BibitemShut {NoStop}%
\bibitem [{\citenamefont {Cochran}\ \emph {et~al.}(2024)\citenamefont
  {Cochran}, \citenamefont {Jobst}, \citenamefont {Rosenberg}, \citenamefont
  {Lensky}, \citenamefont {Gyawali} \emph {et~al.}}]{cochran2024visualizing}%
  \BibitemOpen
  \bibfield  {author} {\bibinfo {author} {\bibfnamefont {T.~A.}\ \bibnamefont
  {Cochran}}, \bibinfo {author} {\bibfnamefont {B.}~\bibnamefont {Jobst}},
  \bibinfo {author} {\bibfnamefont {E.}~\bibnamefont {Rosenberg}}, \bibinfo
  {author} {\bibfnamefont {Y.~D.}\ \bibnamefont {Lensky}}, \bibinfo {author}
  {\bibfnamefont {G.}~\bibnamefont {Gyawali}}, \emph {et~al.},\ }\href
  {https://arxiv.org/abs/2409.17142} {\bibinfo {title} {Visualizing dynamics of
  charges and strings in (2+1)d lattice gauge theories}} (\bibinfo {year}
  {2024}),\ \Eprint {https://arxiv.org/abs/2409.17142} {arXiv:2409.17142}
  \BibitemShut {NoStop}%
\bibitem [{zen()}]{zenodo}%
  \BibitemOpen
  \href@noop {} {\bibinfo {title} {All data and simulation codes are available
  upon reasonable request at
  \href{https://doi.org/10.5281/zenodo.14794905}{10.5281/zenodo.14794905}}}\BibitemShut
  {NoStop}%
\bibitem [{\citenamefont {Bissbort}(2013)}]{bissbort2013dynamical}%
  \BibitemOpen
  \bibfield  {author} {\bibinfo {author} {\bibfnamefont {U.}~\bibnamefont
  {Bissbort}},\ }\emph {\bibinfo {title} {Dynamical effects and disorder in
  ultracold bosonic matter}},\ \href@noop {} {\bibinfo {type}
  {doctoralthesis}},\ \bibinfo  {school} {Universit{\"a}tsbibliothek Johann
  Christian Senckenberg} (\bibinfo {year} {2013})\BibitemShut {NoStop}%
\end{thebibliography}%

\end{document}